\documentclass[reprint,pra,superscriptaddress,showpacs]{revtex4-1}
\usepackage{hyperref}
\usepackage{bbm,bm}
\usepackage{amsmath, amssymb}
\usepackage{xcolor}
\usepackage{graphicx,epsfig}
\usepackage{pifont}

\begin{document}
	
	\title{Coarse-graining in the derivation of Markovian master equations and its significance in quantum thermodynamics}
	\author{J.\ D.\ Cresser}
	\email{james.cresser@glasgow.ac.uk}
	\affiliation{Department of Physics and Astronomy, University of Glasgow, Glasgow, G12 8QQ UK}
	\affiliation{Department of Physics and Astronomy, Macquarie University, 2109 NSW, Australia}
	\author{C.\ Facer}
	\affiliation{Department of Physics and Astronomy, Macquarie University, 2109 NSW, Australia}
	\affiliation{Displayr Australia Pty Ltd,  Glebe 2037 NSW, Australia}

	\begin{abstract}
		The coarse-graining approach to deriving the quantum Markovian master equation is revisited, with close attention given to the underlying approximations. It is further argued that the time interval over which the coarse-graining is performed is a free parameter that can be given a physical measurement-based interpretation. In the case of the damping of composite systems to reservoirs of different temperatures, currently of much interest in the study of quantum thermal machines with regard to the validity of `local' and `global' forms of these equations, the coupling of the subsystems leads to a further timescale with respect to which the coarse-graining time interval can be chosen. Different choices lead to different master equations that correspond to the local and global forms. These can be then understood as having different physical interpretations based on the role of the coarse-graining, as well as different limitations in application.
	\end{abstract}

	\maketitle
	\section{Introduction}

	The master equation has long been an essential tool in describing the dynamics of the reduced density operator of an open quantum system. This is particularly the case for systems described by Markovian master equations, equations which can be informally understood as describing a dynamic which possesses `no memory'. The  master equation is known in this case to take a particular form \cite{Lindblad76, GKS76}, typically referred to as the Lindblad form. However there are on-going issues that arise where the derivation of these equations from a microscopic model is concerned.  
	
	Physically, a master equation describes the dynamics of a system coupled to some other, usually much larger system, typically the environment with which the system is invariably in interaction, and this usually modelled as a thermodynamic reservoir, and as such, the master equation is derived from a microscopic description of the system+reservoir. Except in a small handful of instances, of which an important example is that of harmonic oscillators coupled to a reservoir modelled as a collection of harmonic oscillators, the derivation of the exact master equation proves to be an intractable problem. However, fortunately, for many systems of on-going interest, approximations can be invoked yielding a master equation that is of the required Lindblad form. 
	
	There are a number of ways that this master equation can be derived, though all of the derivations rely on there being a wide separation of typical timescales making up the dynamics, these timescales typically being the correlation time of the reservoir $\tau_c$, the system evolution timescale $\tau_S$, typically a system decay time, and system Bohr frequency timescales $\omega_S^{-1}$, with $\tau_c\ll\omega_S^{-1}\ll\tau_S$. One way or another, a coarse-graining, or smoothing of this dynamics over the very short time scales $\tau_c$ and $\omega_S^{-1}$ plays an essential role in the derivation. In the early work of Bloch and Redfield \cite{Bloch57,Redfield57} and later Cohen-Tannoudji \emph{et.\ al}.\ \cite{Cohen86,Cohen92}, the coarse-graining is initially done explicitly through the introduction of a coarse-graining interval $\Delta t$ which is constrained to satisfy $\tau_c\ll\Delta t\ll\tau_S$. But in \cite{Bloch57,Redfield57}, the role of explicit coarse-graining is bypassed and Born and Markov approximations introduced that nowadays are commonly implemented in derivations such as that presented in \cite{BrPet} where coarse-graining is not made explicit, though arguments based on timescales are used to introduce these approximations to simplify the exact master equation to a tractable form. Typically, a further approximation is required, the secular approximation, which can be understood as an averaging over terms in the master equation that are rapidly oscillating with a frequency $\sim\omega_S^{-1}$, and is hence also a timescale based argument, leading to the required Lindblad form. 
	
	Variations on the explicit coarse-graining approach have begun to appear increasingly often \cite{Bacon99,Lidar01,Kryszewski08,Schaller08,Benatti09,Benatti10,Majenz13, Facer14,Buchheit16}. There are advantages in the latter method. Apart from anything else, the explicit coarse-graining method leads directly to a master equation of Lindblad form for essentially any choice of $\Delta t$, i.e., the secular approximation is not required as a separate step, though the secular approximated form of the master equation will also follow for $\Delta t$ made sufficiently large \cite{Cohen86,Cohen92}. Being able to choose $\Delta t$ in such a way emphasizes the point that the coarse-graining time interval over which the dynamics are smoothed is in fact a free parameter of the theory. Different choices of $\Delta t$ can be made, with an emphasis on, for instance, obtaining better approximations to the exact master equation \cite{Schaller08}, or else aiming to consistently include interference (or cross-damping) terms \cite{Buchheit16}. 

	  The choice of time scale acquires greater significance when the evolution of the system has more than one time scale beyond the typical ones listed above. This situation is encountered in general when the system structure is such that there are dynamics internal to the system with their own well-defined time scale such as, for instance, for composite systems in which the internal coupling of the systems gives rise to a further time scale $\Omega^{-1}$ where $\tau_c\ll\omega_S^{-1}\ll\Omega^{-1}\ll\tau_S$. Such composite systems are currently of considerable interest in quantum thermodynamical applications, particularly when the individual subsystems are part of some kind of heat engine, and where each subsystem is coupled to separate thermodynamic reservoirs at different temperatures so that heat can flow from one reservoir to the other. In such instances, the question arises as to how to model the damping of the system(s), whether by the local or global approaches to deriving the Born-Markov-secular approximated form of the master equation \cite{Hofer17}. Local coupling describes a model in which each reservoir is coupled only to the energy eigenstates of its associated subsystem, while global coupling describes a model in which each reservoir is coupled to the energy eigenstate of the combined system. The former has the unwanted property of predicting, for reservoirs all of the same temperature (or only one reservoir) a steady state which is not the expected Gibbs distribution, which can only be resolved by using the global approach, an issue which was first noted in the work of Carmichael and Walls \cite{Carmichael73} for coupled, damped harmonic oscillators, and subsequently dealt with for the damped Jaynes-Cummings model in \cite{Cresser92}. But the global approach can lead to unexpected outcomes, e.g., \cite{Wichterich07} examine a circumstance in which there is no heat flow at all between reservoirs of different temperatures if the secular approximate forms of the master equation is used. The presence of an extra internal time scale also raises the possibility of not making the secular approximation (achieved by coarse-graining on a timescale $\Delta t\gg \Omega^{-1}$), but by taking $\Delta t\ll\Omega^{-1}$, referred to here as a partial secular approximation which will lead to a different but still Lindbladian master equations for the same set-up, but staying within the global approach. The impact of this choice on the expected heat flow between reservoirs of different temperatures if the appropriate master equation is used is one of the issues addressed in this paper.
	 
	  It can further be argued that $\Delta t$ can have a physical interpretation in terms of the temporal resolution of measurements made on the environment. The physical interpretations of the different forms of the master equation, apart from the fact that they are different approximations to the same underlying exact dynamics, is also investigated in this paper. 
	  
	  The paper is structured as follows. In Section \ref{MEDerivation}, the master equation derivation is presented in a way that leads to both the usual Bloch-Redfield result, and the coarse-grained version of \cite{Cohen92}. Some comments on the Markov approximation, and a collisional model interpretation of the coarse-grained result, are also presented. In Section \ref{Coupling}, the particular case of a system coupled to a bosonic reservoir is analysed, and the limiting forms for the full secular approximation and a partial secular aproximation for composite systems are introduced along with a proof that the partial secular approximated form for the master equation is still of Lindblad form. A measurement interpretation of the coarse-grained master equation is discussed in Section \ref{MeasInterp} and the generalisation of some preceding results to the case of more than one independent reservoir is given in Section \ref{Section:MoreThanOneReservoir}. Examples of composite systems are then given in Section \ref{Examples}.

\section{Master equation derivation\label{MEDerivation}}

Given a microscopic model of a system $S$ interacting with a reservoir $R$ as described by the total Hamiltonian of the combined system $S\oplus R$:
\begin{equation}
	H=H_S+H_R+V=H_0+V
	\label{Hamiltonian}
\end{equation}
the evolved combined system density operator $\chi(t)$ is then given by $\chi(t)=U(t)\chi(0)U^\dagger(t)$ where $U(t)=\exp(-iHt/\hbar)$ and the system density operator at time $t$ given by $\rho(t)=\text{Tr}_R[\chi(t)]$.  We will work in the interaction picture (indicated by overbars) so that the state of the combined system plus reservoir will evolve according to
\begin{equation}
	\bar{\chi}(t)=\bar{U}(t,0)\chi(0)\bar{U}^\dagger(t,0)
\end{equation}
where the time evolution operator in the interaction picture 
\begin{equation}
	\bar{U}(t,0)=e^{iH_0t/\hbar}e^{-iHt/\hbar}
\end{equation}
satisfies
\begin{equation}
	\frac{d\bar{U}}{dt}=-\frac{i}{\hbar}\bar{V}(t)\bar{U}
\end{equation}
with
\begin{equation}
	\bar{V}(t)=e^{iH_0t/\hbar}Ve^{-iH_0t/\hbar}
\end{equation}
and the initial state must be taken to be the product state, $\chi(0)=\rho(0)\otimes\rho_R(0)$.

A refined version of the arguments of \cite{Cohen92} are presented below, leading to a second order perturbative result that yields, in different limits, both the required coarse-grained master equation, as well as the Born-Markov master equation.

\subsection{Perturbative expansion of system density operator}

The derivation begins by obtaining an exact expression for the evolution of the system density operator over an interval $(t,t+\Delta t)$
\begin{align}
	\bar{\rho}(t+\Delta t)=&~\text{Tr}_R
	\left[\bar{\chi}(t+\Delta t)\right]\notag\\
	=&~\text{Tr}_R\left[\bar{U}(t+\Delta t,t)\bar{\chi}(t)
	\bar{U}^\dagger(t+\Delta t,t)\right]
\end{align}
where $\bar{U}(t+\Delta t,t)=\bar{U}(t,0)\bar{U}(t+\Delta t,0)^{-1}$. We now note, as is done in Cohen-Tannoudji \cite{Cohen92}, that $\bar{\chi}(t)$ can be written $\bar{\chi}(t)=\bar{\rho}(t)\otimes\bar{\rho}_R(t)+\left[\bar{\chi}(t)-\bar{\rho}(t)\otimes\bar{\rho}_R(t)\right]$ where the second term represents the contribution to $\bar{\chi}(t)$ due to the correlations between the system and the reservoir at time $t$. But we can proceed further and separate out the corrections arising due to making the Born approximation. In this approximation it is assumed that, as far as the system is concerned, the state of the reservoir can be taken as being unchanged by the interaction, and further that any correlations that develop between the system and the reservoir over the time interval $(0,t)$ are ignored. This can be implemented here by writing
\begin{align}
	\bar{\chi}(t)=&~\bar{\rho}(t)\otimes\rho_R(0)
	+\bar{\rho}(t)\otimes\left[\rho_R(t)-\rho_R(0)\right]\notag\\
	&+\left[\bar{\chi}(t)-\bar{\rho}(t)\otimes\bar{\rho}_R(t)\right]\notag\\
	=&~\bar{\rho}(t)\otimes\rho_R(0)+\bar{\chi}_{corr}(t)
	\label{CorrectionsExtracted}
\end{align}
in which $\bar{\chi}_{corr}(t)$ is the contributions due to the correlations that develop between the system and the reservoir up to time $t$, \emph{plus} corrections associated with changes to the state of the reservoir due to its coupling to the system, usually ignored under the Born approximation. Note that $\bar{\chi}_{corr}$ does not give rise to a `correction' to the system density operator since $\text{Tr}_R\left[\bar{\chi}_{corr}(t)\right]=0$.

If we now consider evolution for a further period $\Delta t$, we could then trivially write $\bar{\chi}(t+\Delta t)=\bar{\rho}(t+\Delta t)\otimes\rho_R(0)+\bar{\chi}_{corr}(t+\Delta t)$, but this is not of much value. Instead, we can also write
\begin{align}
	\bar{\chi}(t+\Delta t)=&~\bar{U}(t+\Delta t,t)\bar{\rho}(t)\otimes\rho_R(0)\bar{U}^\dagger(t+\Delta t,t)\notag\\
	&+\bar{U}(t+\Delta t,t)\bar{\chi}_{corr}(t)\bar{U}^\dagger(t+\Delta t,t)
\end{align}
so that
\begin{align}
	\bar{\rho}(t+\Delta t)=&~\text{Tr}_R\left[\bar{U}(t+\Delta t,t)\bar{\rho}(t)\otimes\rho_R(0)\bar{U}^\dagger(t+\Delta t,t)\right]\notag\\
	&+\text{Tr}_R\left[\bar{U}(t+\Delta t,t)\bar{\chi}_{corr}(t)\bar{U}^\dagger(t+\Delta t,t)\right].
	\label{rhotDt}
\end{align}
The second term on the right hand side, which is now non-zero for $\Delta t\ne 0$, is the contribution to the state of the system at time $\bar{\rho}(t+\Delta t)$ due to the correlations/reservoir state present at time $t$. But it also represents an error that accumulates over the time interval $(t,t+\Delta t)$ if the exact density operator at time $t$ is approximated by the product state  $\bar{\rho}(t)\otimes\rho_R(0)$.

We now want to expand this expression to second order in the interaction $V$. In doing so, the separation given by Eq.~(\ref{CorrectionsExtracted}) makes it possible to implement the machinery of the Zwanzig-Nakajima projection operator method \cite{Nakajima58,Zwanzig60}, much used in deriving formally exact non-Markovian master equations. We first introduce the projection operators $\mathcal{P},\mathcal{Q}=1-\mathcal{P}$ with $\mathcal{P}$ defined by
\begin{equation}
	\mathcal{P}\bar{\chi}(t)=\text{Tr}_R\left[\bar{\chi}(t)\right]\otimes\rho_R(0)=\bar{\rho}(t)\otimes\rho_R(0).
\end{equation}
We then find that $\bar{\chi}_{corr}(t)=\mathcal{Q}[\bar{\chi}(t)]$ so that, as already expected, $\text{Tr}_R[\bar{\chi}_{corr}(t)]=\text{Tr}_R[\mathcal{Q}[\bar{\chi}(t)]]=0$. We then have, from Eq.~(\ref{rhotDt})
\begin{align}
	\bar{\rho}(t+\Delta t)
	=&~\text{Tr}_R\left[\bar{U}(t+\Delta t,t)
	\bar{\rho}(t)\otimes\rho_R(0)\bar{U}^\dagger(t+\Delta t,t)\right]\notag\\
	&+\text{Tr}_R\left[\bar{U}(t+\Delta t,t)\mathcal{Q}
	\left[\bar{\chi}(t)\right]\bar{U}^\dagger(t+\Delta t,t)\right].
	\label{ExactDeltarho}
\end{align}
The usual procedure for deriving the equations of motion of $\mathcal{P}\left[\bar{\chi}\right]$ and $\mathcal{Q}\left[\bar{\chi}\right]$ can be employed here. The equation of motion for $\bar{\chi}(t)$
\begin{equation}
	\frac{d\bar{\chi}}{dt}=-\frac{i}{\hbar}\left[\bar{V}(t),\bar{\chi}\right]
	=\bar{\mathcal{L}}(t)\bar{\chi}
	\label{LDefined}
\end{equation}
has the formal solution $\bar{\chi}(t)=\bar{U}(t,0)\bar{\chi}(0)\bar{U}^\dagger(t,0)
=\bar{\mathcal{G}}(t,0)\bar{\chi}(0)$ where the free propagator for the system plus reservoir in the interaction picture is given by
\begin{equation}
	\bar{\mathcal{G}}(t,t')=T\exp\left(\int_{t'}^{t}\bar{\mathcal{L}}(t'')dt''\right)
\end{equation}
where $T$ indicates time ordering. This yields the following equation of motion for $\mathcal{Q}[\bar{\chi}(t)]$:
\begin{equation}
	\frac{d\mathcal{Q}[\bar{\chi}(t)]}{dt}=\mathcal{Q}\bar{\mathcal{L}}(t)\mathcal{Q}[\bar{\chi}(t)]+\mathcal{Q}\bar{\mathcal{L}}(t)\mathcal{P}[\bar{\chi}(t)].
\end{equation}
On using the initial condition $\mathcal{Q}[\bar{\chi}(0)]=\mathcal{Q}[\rho(0)\otimes\rho_R(0)]=0$ the formal solution for $\mathcal{Q}\left[\bar{\chi}(t)\right]$ is
\begin{equation}
	\mathcal{Q}\left[\bar{\chi}(t)\right]
	=\int_{0}^{t}\bar{\mathcal{G}}_\mathcal{Q}(t,t_1)
	\mathcal{Q}\left[\bar{\mathcal{L}}(t_1)\mathcal{P}\left[\bar{\chi}(t_1)\right]\right]dt_1
\end{equation}
where
\begin{equation}
	\bar{\mathcal{G}}_\mathcal{Q}(t,t')
	=T\exp\left(\int_{t'}^{t}\mathcal{Q}\bar{\mathcal{L}}(t'')dt''\right).
\end{equation}
\begin{widetext}
	In terms of these propagators we can then write 
\begin{align}
	\bar{\rho}(t+\Delta t)=&~\text{Tr}_R\left[\bar{\mathcal{G}}(t+\Delta t,t)\mathcal{P}\left[\bar{\chi}(t)\right]\right]\notag\\
	&+\text{Tr}_R\left[\bar{\mathcal{G}}(t+\Delta t,t)\int_{0}^{t}dt_1\bar{\mathcal{G}}_\mathcal{Q}(t,t_1)\mathcal{Q}\left[\bar{\mathcal{L}}(t_1)\mathcal{P}\left[\bar{\chi}(t_1)\right]\right]\right]
\end{align}
We are only going to consider contributions to second order in the interaction $V$, so we will make the approximations
\begin{equation}
	\bar{\mathcal{G}}(t+\Delta t,t)\approx
	1+\int_{t}^{t+\Delta t}dt_1\bar{\mathcal{L}}(t_1)+\int_{t}^{t+\Delta t}dt_2\int_{t}^{t_2}dt_1\bar{\mathcal{L}}(t_2)\bar{\mathcal{L}}(t_1)
	\label{GBar2ndOrder}
\end{equation}
and
\begin{equation}
	\bar{\mathcal{G}}_\mathcal{Q}(t,t_1)\approx 1+\int_{t_1}^{t}dt''\mathcal{Q}\,\bar{\mathcal{L}}(t'').
\end{equation}
Making use of the exact results $\text{Tr}_R\left[\bar{\mathcal{L}}(t)\rho_R(0)\right]=0$ and $\text{Tr}_R[\mathcal{Q}[\chi]]=0$ as well as Eq.\ (\ref{LDefined}) then gives
\begin{equation}
	\begin{aligned}
		\frac{\bar{\rho}(t+\Delta t)-\bar{\rho}(t)}{\Delta t}
		\approx&~-\frac{1}{\hbar^2}\frac{1}{\Delta t}
		\int_{t}^{t+\Delta t}dt_2\int_{t}^{t_2}dt_1
		\text{Tr}_R\left[\left[\bar{V}(t_2),\left[\bar{V}(t_1),\bar{\rho}(t)\otimes\rho_R(0)\right]\right]\right]\\
		&~-\frac{1}{\hbar^2}\frac{1}{\Delta t}
		\int_{t}^{t+\Delta t}dt_2\int_{0}^{t}dt_1
		\text{Tr}_R\left[\left[\bar{V}(t_2),\left[\bar{V}(t_1),\bar{\rho}(t_1)
		\otimes\rho_R(0)\right]\right]\right]
	\end{aligned}
	\label{deltarho}
\end{equation}
This result is similar to that found in \cite{Cohen92}, Eq.\ (IV~D.7), though there the separation Eq.\ (\ref{CorrectionsExtracted}) was not used, and a phenomenological argument was used to determine the form of the second term. There are now two directions that this result can be taken.
\end{widetext}

\subsubsection{No coarse-graining limit}

There is nothing in the above derivation that places, at this stage, any limit on $\Delta t$ other than being sufficiently small that a second order expansion in the interaction $V$ is valid. In particular, we can now let $\Delta t\to 0$, i.e., implying that there is no coarse-graining. The left hand side is just the derivative of $\bar{\rho}$, the first term on the right hand side of Eq.~(\ref{deltarho}) vanishes and the second reduces to
\begin{equation}
	\frac{d\bar{\rho}}{dt}=-\frac{1}{\hbar^2}\int_{0}^{t}dt_1
	\text{Tr}_R
	\left[\big[\bar{V}(t),[\bar{V}(t_1),\bar{\rho}(t_1)\otimes\rho_R(0)]\big]\right].
	\label{PreMarkovAgain}
\end{equation}
This is the post-Born approximation result obtained in a typical derivation of the Bloch-Redfield master equation \cite{BrPet}. There is no explicit coarse-graining in time. The implicit coarse-graining of the subsequent Markov approximation in which $\rho(t_1)\approx\rho(t)$ then yields
\begin{equation}
	\frac{d\bar{\rho}}{dt}=-\frac{1}{\hbar^2}\int_{0}^{\infty}dt_1
	\text{Tr}_R
	\left[\big[\bar{V}(t),[\bar{V}(t_1),\bar{\rho}(t)\otimes\rho_R(0)]\big]\right]
\end{equation}
with, if necessary, the secular approximation to follow to then yield the final Lindblad form of the master equation.

\subsubsection{Coarse-graining limit}

An alternate derivation as exemplified by that of Cohen-Tannoudji \cite{Cohen86, Cohen92} makes the coarse graining time scale $\Delta t$ an explict part of the derivation of the master equation, and amounts to approximating the derivative of the interaction picture density operator by a coarse-grained version.
\begin{equation}
	\frac{d\bar{\rho}}{dt}\approx\frac{\Delta\bar{\rho}}{\Delta t}=\frac{\bar{\rho}(t+\Delta t)-\bar{\rho}(t)}{\Delta t}
	\label{CoarseGrainingDefinition}
\end{equation}
i.e., the instantaneous rate of change $d\bar{\rho}/dt$ is effectively smoothed out on a time scale $\Delta t$ -- the `coarse-grained' time scale -- where $\tau_c\ll\Delta t\ll\tau_S$. Coarse-graining in this manner was the starting point in the early work of Redfield \cite{Redfield57}, but close attention to some of the underlying approximations was first done in \cite{Cohen92} where it was shown that this coarse-grained derivative emerges naturally under conditions in which correlations embodied in $\bar{\chi}_{corr}$ that develop between the system and the reservoir are negligible, and the Born approximation is valid.  

The explicit coarse-graining result follows if $\Delta t$ is kept non-zero, and $t$ and $\Delta t$ chosen such that $t,\Delta t\gg\tau_c$. The arguments of \cite{Cohen92} can then be applied. The first term in Eq.\ (\ref{deltarho}) will be of order $1/\tau_S$, while the second term, by virtue of the fact that the integrals are over non-overlapping intervals, will be of order $(1/\tau_S)\cdot(\tau_c/\Delta t)$ and hence can be ignored relative to the first. The argument is outlined further in Appendix \ref{TimeScales}. This result now provides us with an approximate form for the coarse-grained time derivative

\begin{widetext}
	
\begin{equation}
	\frac{\Delta\bar{\rho}}{\Delta t}\approx-\frac{1}{\hbar^2}\frac{1}{\Delta t}
	\int_{t}^{t+\Delta t}dt_2\int_{t}^{t_2}dt_1
	\text{Tr}_R\left[\big[\bar{V}(t_2),[\bar{V}(t_1),\bar{\rho}(t)\otimes\rho_R(0)]\big]\right].
	\label{CTEqn}
\end{equation}
The required coarse-grained master equation in the Schr\"{o}dinger picture is then obtained by noting that 
\begin{equation}
	\frac{d\rho}{dt}=-\frac{i}{\hbar}\left[H_0,\rho\right]+e^{-iH_0t/\hbar}
	\frac{d\bar{\rho}}{dt}e^{iH_0t/\hbar}.
\end{equation}
and we now approximate the exact time derivative on the right hand side by its coarse-grained approximation, in accordance with Eq.~(\ref{CoarseGrainingDefinition}), giving
\begin{align}
	\frac{d\rho}{dt}\approx&~
	-\frac{i}{\hbar}\left[H_0,\rho\right]+e^{-iH_0t/\hbar}
	\frac{\Delta\bar{\rho}}{\Delta t}e^{iH_0t/\hbar}\notag\\
	\approx&~-\frac{i}{\hbar}\left[H_0,\rho\right]
	-\frac{1}{\hbar^2}\frac{1}{\Delta t}
	\int_{t}^{t+\Delta t}dt_2\int_{t}^{t_2}dt_1
	\text{Tr}_R\left[\big[\bar{V}(t_2-t),[\bar{V}(t_1-t),\rho(t)\otimes\rho_R(0)]\big]\right].
\end{align}
A simple change of variable then yields the approximate result
\begin{equation}
	\frac{d\rho}{dt}\approx-\frac{i}{\hbar}\left[H_0,\rho\right]
	-\frac{1}{\hbar^2}\frac{1}{\Delta t}
	\int_{0}^{\Delta t}dt_2\int_{0}^{t_2}dt_1
	\text{Tr}_R\left[\big[\bar{V}(t_2),[\bar{V}(t_1),\rho(t)\otimes\rho_R(0)]\big]\right].
\end{equation}
This is an approximate equation for the exact density operator $\rho(t)$. We can, instead, write an exact equation for the approximate density operator, which we could call, for instance, $\rho_{\Delta t}(t)$ to emphasize that it has been obtained by coarse-graining over a time interval $\Delta t$. But, instead of using this cumbersome notation, we will instead simply continue to write $\rho(t)$, which then satisfies the equation
\begin{equation}
	\frac{d\rho}{dt}=-\frac{i}{\hbar}\left[H_0,\rho\right]
	-\frac{1}{\hbar^2}\frac{1}{\Delta t}
	\int_{0}^{\Delta t}dt_2\int_{0}^{t_2}dt_1
	\text{Tr}_R\left[\big[\bar{V}(t_2),[\bar{V}(t_1),\rho(t)
	\otimes\rho_R(0)]\big]\right]
	\label{OrigCTEqn}
\end{equation}
with the understanding that solving this equation will yield a coarse-grained approximate result for the exact $\rho$. This is the coarse-grained master equation as given by Cohen-Tannoudji \cite{Cohen92}, Eq.\ (IV B.30), and derived elsewhere by similar arguments.

As has been shown in a number of places \cite{Schaller08, Stokes12, Facer14, Buchheit16}, manipulation of the double time integral enables this result to be rewritten as
\begin{equation}
	\frac{d\rho}{dt}=-\frac{i}{\hbar}\left[H_0+\Delta H,\rho\right]
	-\frac{1}{2\hbar^2}\frac{1}{\Delta t}
	\int_{0}^{\Delta t}dt_2\int_{0}^{\Delta t}dt_1
	\text{Tr}_R\Big[\big[\bar{V}(t_2),[\bar{V}(t_1),\rho(t)
	\otimes\rho_R(0)]\big]\Big]
	\label{MyEquation}
\end{equation}
\end{widetext}
with the energy shift $\Delta H$ given by
\begin{equation}
	\Delta H=-\frac{i}{2\hbar}\frac{1}{\Delta t}
	\int_{0}^{\Delta t}dt_2\int_{0}^{t_2}dt_1
	\text{Tr}_R\left[[\bar{V}(t_2),\bar{V}(t_1)]\rho_R(0)\right].
	\label{CoherentPart}
\end{equation}
Hereinafter, the energy shift term will assumed to be sufficiently small as to be ignored.

Defining
\begin{equation}
	\mathcal{V}(t_2,t_1)=\int_{t_1}^{t_2}\bar{V}(t)dt
\end{equation}
the dissipator in Eq.\ (\ref{MyEquation}) becomes
\begin{equation}
	\mathcal{D}[\rho]=-\frac{1}{2\hbar^2}\frac{1}{\Delta t}\text{Tr}_R\left[\left[\mathcal{V}(\Delta t,0),\left[\mathcal{V}(\Delta t,0),\rho(t)\otimes\rho_R(0)\right]\right]\right]
	\label{CoarseGrainedLindblad}
\end{equation}
which is clearly of the required Lindblad form. The secular approximation as a separate step in the calculation is not needed to achieve this. Nevertheless, the secular approximated master equation can be regained here by a suitably large choice of $\Delta t$, but the generality of the result Eq.\ (\ref{CoarseGrainedLindblad}) implies that master equations not of the secular approximated form, but nevertheless still Lindblad, can be obtained, as discussed below in Section \ref{Coupling}.

\subsection{Where is the Markov approximation?}

The general result Eq.~(\ref{deltarho}) leads to the two different forms for the master equation. In the first, Eq.~(\ref{PreMarkovAgain}), the Markov approximation is yet to be implemented, and would be done so in the usual way \cite{BrPet}. In the second, equation Eq.\ (\ref{MyEquation}) is clearly Markovian, but the Born and Markov approximations have emerged as a consequence of the substitution Eq.~(\ref{CorrectionsExtracted}), showing that \emph{provided the coarse-graining time $\Delta t$ is made sufficiently large}, the memory contributions implied by the correction term $\bar{\chi}_{corr}$ can be neglected. Thus, the Markov nature of the master equation has emerged as a consequence of the coarse-grained averaging which has smoothed out the evolution of $\rho(t)$ on a coarse-grained time-scale $\Delta t$.

\subsection{Collisional model interpretation}

Returning to the expression Eq.\ (\ref{ExactDeltarho}) in the $\Delta t\gg\tau_c$ limit, which can be written, with $t=n\Delta t$, $\bar{\rho}_n=\bar{\rho}(n\Delta t)$, and $\bar{\mathcal{G}}_n=\bar{\mathcal{G}}(n\Delta t,(n-1)\Delta t)$ as
\begin{equation}
	\bar{\rho}_{n+1}\otimes\rho_R(0)=\mathcal{P}\mathcal{G}_{n+1}\bar{\rho}_n\otimes\rho_R(0)
\end{equation}
which can be iterated and after tracing over the reservoir states gives
\begin{equation}
	\bar{\rho}_{n}=\text{Tr}_R\left[\mathcal{G}_n\mathcal{P}\mathcal{G}_{n-1}
	\ldots \mathcal{P}\mathcal{G}_1[\rho(0)\otimes\rho_R(0)]\right].
\end{equation}
This result has a ready interpretation. The system is exposed to the reservoir for an interval of time $\Delta t$ after which interaction ceases, and the reservoir returned to its original state $\rho_R(0)$. Thereafter the system and reservoir in its initial state $\rho_R(0)$ come into interaction again for a further period $\Delta t$ and so on. This is a picture reminiscent of the quantum collision models of open quantum systems currently attracting considerable attention, see e.g., \cite{Vacchini16,Kretschmer16} and references therein.

\section{Coupling to a bosonic thermal reservoir\label{Coupling}}

We will now use the above result to look at the often encountered example of a system coupled to a reservoir of simple harmonic oscillators for which the interaction is
\begin{equation}
	V=BX
	\label{BathSystemCoupling}
\end{equation} 
where $X$ is some system operator and $B$ is the reservoir operator
\begin{equation}
	\bar{B}(t)=\int_{0}^{\infty}g(\omega)\left(b(\omega)e^{-i\omega t}+b^\dagger(\omega)e^{-i\omega t}\right)d\omega.
	\label{ReservoirOperator}
\end{equation}
where $\left[b(\omega),b^\dagger(\omega')\right]=\delta(\omega-\omega')$ and with
\begin{equation}
	H_R=\int_{0}^{\infty}\hbar\omega b^\dagger(\omega)b(\omega)\,d\omega.
\end{equation}
The above choice of model for the reservoir is much like that for the quantized electromagnetic field, and so excitations of the reservoir will be referred to below as photons, though it would probably be more appropriate to label them simply as `quanta'.
 
The reservoir is assumed to be in the thermal equilibrium state
\begin{equation}
	\rho_R(0)=\frac{e^{-H_R/kT}}{\text{Tr}_R[e^{-H_R/kT}]}.
	\label{ThermalEquilState}
\end{equation}
\begin{widetext}
	The dissipative part of the master equation Eq.\ (\ref{MyEquation}) can then be written
\begin{equation}
	\mathcal{D}[\rho]
	=\frac{1}{\hbar^2\Delta t}
	\int_{0}^{\Delta t}dt_2
	\int_{0}^{\Delta t}dt_1G(t_2-t_1)\left[\bar{X}(t_1)\rho \bar{X}(t_2)
	-\tfrac{1}{2}\left\{\bar{X}(t_2)\bar{X}(t_1),\rho\right\}\right].
	\label{GeneralDissipator}
\end{equation}
where the reservoir correlation function is
\begin{align}
	G(t_2-t_1)=&~\text{Tr}_R\left[B(t_2)B(t_1)\rho_R\right]\notag\\
	=&~\int_{0}^{\infty}g(\omega)^2\left(
		(2n(\omega)+1)\cos\omega (t_2-t_1)-i\sin\omega (t_2-t_1)\right)\,d\omega,
	\label{Gkernel}
\end{align}
\end{widetext}
where the reservoir occupation number at frequency $\omega$ is given by the Planck formula
\begin{equation}
	n(\omega)=\left(e^{\hbar\omega/kT}-1\right)^{-1}
\end{equation}
and where $G(t)$ has the symmetry property $G^*(t)=G(-t)$.

The spectral density $g(\omega)^2$ will be assumed to be Ohmic with a high frequency cutoff such that the correlation function $G(t)$ will decay on a time scale $\tau_c$. For the reservoir at zero temperature, $\tau_c$ will be the vaccuum correlation decay time $\tau_v$. As usual, $g(\omega)$ will be assumed to be a sufficiently broad function that $\tau_v$ will be very short compared to the time scale $\tau_S$ characterizing the evolution of the system. For non-zero temperatures the correlation time $\tau_c$ will be determined by both $\tau_v$ and the correlation time of the thermal reservoir, $\tau_T\sim\hbar/2\pi kT$, with $\tau_v$ dominating at high temperatures when $\tau_T\ll\tau_v$ and at low temperatures, but otherwise $\tau_c\sim\tau_T$. This time scale $\tau_c$ will in all instances be assumed to be very short compared to time scales characterizing the evolution of the system, $\tau_v,\tau_c\ll\omega_S^{-1}\ll\tau_S$ where $\omega_S$ is a typical Bohr frequency for the system. A more detailed analysis is to be found in \cite{Rivas10}, based on the work of Carmichael \cite{Carmichael99} who also gives typical estimates of these time scales for optical systems, for which the inequalities are well-satisfied.

Simplification of this result can be made in a standard way by expanding the system operator $\bar{X}(t)$ in the energy basis of the unperturbed Hamiltonian $H_0$:
\begin{equation}
	\bar{X}(t)=e^{iH_0t/\hbar}Xe^{-iH_0t/\hbar}
	=\sum_{a,b}^{}|a\rangle\langle b|\langle a|X|b\rangle e^{i(\omega_a-\omega_b)t}.
\end{equation}
Now introduce the eigenoperators
$$X_m=\sum_{a,b}^{}|a\rangle\langle b|\langle a|X|b\rangle\delta_{\omega_m,\omega_a-\omega_b}$$
which are such that $\left[H_0,X_m\right]=\hbar\omega_mX_m$. By construction, if $m\ne n$, then $\omega_m\ne\omega_n$. In fact, we can order the frequencies such that $\omega_m>\omega_n$ if $m>n$, and we can further adopt the convention that $\omega_n\lessgtr 0$ if $n\lessgtr 0$ and $\omega_0=0$.

The expression for $\bar{X}(t)$ can now be written
\begin{equation}
	\bar{X}(t)=\sum_{m=-N}^{N}X_me^{i\omega_mt}
	=\sum_{m=-N}^{N}X^\dagger_me^{-i\omega_mt}
\end{equation}
from which follows that $X^\dagger_m=X_{-m}$ and $\omega_m=\omega_{-m}$.

The master equation now becomes, ignoring energy level shifts,
\begin{align}
	\frac{d\rho}{dt}=-\frac{i}{\hbar}\left[H_S,\rho\right]+\sum_{m,n}^{}
	\gamma_{mn}(\Delta t)
	\left[X_m\rho X_n^\dagger-\frac{1}{2}\{X_n^\dagger X_m,\rho\}\right]
	\label{TheBXMasterEquation}
\end{align}
where the matrix elements $\gamma_{mn}(\Delta t)$ can be expressed as
\begin{equation}
	\gamma_{mn}(\Delta t)=\frac{1}{\hbar^2\Delta t}
	\int_{0}^{\Delta t}dt_2\int_{0}^{\Delta t}dt_1
	G(t_2-t_1)e^{i\omega_m t_1}e^{-i\omega_n t_2}.
\end{equation}
From this expression, and from $G(t)=G^*(-t)$, we immediately see that $\gamma_{mn}(\Delta t)=\gamma_{nm}^*(\Delta t)$ so that the $\gamma_{mn}(\Delta t)$ are elements of a Hermitean matrix which can be readily shown to be positive semidefinite, as expected from the general result Eq.\ (\ref{GeneralLindblad}). Thus we have a legitimate Lindblad equation for any value of $\Delta t$ satisfying $\tau_S\gg\Delta t\gg\tau_c$. However, it is only in a couple of physically distinct limiting cases that, for clear physical reasons, these elements become effectively insensitive to the value of $\Delta t$: the secular approximation limit, and in the circumstance in which the system is a \emph{composite system} for which there are internal time scales for the interaction of the various component parts of the system.

\subsection{Limiting cases for $\gamma_{mn}$}

To arrive at the required expressions for $\gamma_{mn}$, we make the change of variable $t=t_2+t_1$ and $\tau=t_2-t_1$. The expression for $\gamma_{mn}(\Delta t)$ becomes
\begin{widetext}
\begin{align}
	\gamma_{mn}(\Delta t)=&~
	\frac{1}{\hbar^2}\int_{0}^{\Delta t}d\tau
	\left[G(-\tau)e^{i(\omega_m+\omega_n)\tau/2}+G(\tau)e^{-i(\omega_m+\omega_n)\tau/2}\right]\frac{e^{i\omega_{mn}\Delta t}e^{-i\omega_{mn}\tau/2}-e^{i\omega_{mn}\tau/2}}{i\omega_{mn}\Delta t}
	\label{GmnExact}
\end{align}
This result assumes a simple form when $\left\lvert \omega_{mn}\right\rvert\Delta t\ll1$, (which includes $\omega_{mn}=0$) and when $\left\lvert \omega_{mn}\right\rvert\Delta t\gg1$. 

For a given value of $\Delta t$ and for any $\omega_{mn}$ such that $|\omega_{mn}|\Delta t\ll1$, and given that $\Delta t\gg\tau_c$ so that the upper limit on the integral can be taken to be $+\infty$, then we can approximate Eq.~(\ref{GmnExact}) by
\begin{align}
	\gamma_{mn}(\Delta t)=&~\frac{1}{\hbar^2}\int_{-\infty}^{+\infty}G(\tau)e^{-i(\omega_{m}+\omega_n)\tau/2}d\tau\notag\\
	&-\frac{1}{\Delta t}\frac{1}{\hbar^2}\int_{0}^{\infty}
	\left[G(-\tau)e^{i(\omega_m+\omega_n)\tau/2}+G(\tau)e^{-i(\omega_m+\omega_n)\tau/2}\right]\tau d\tau
	\label{GmnApprox}
\end{align}
The first term in Eq.\ (\ref{GmnApprox}) will turn out to be the dominant term and will define the long term evolution timescale $\tau_S$ while, as is shown in Appendix \ref{TimeScales}, the second term in Eq.\ (\ref{GmnApprox}) is $\sim\tau_c/(\Delta t\,\tau_S)$. Thus we have $\gamma_{mn}(\Delta t)\sim \tau_S^{-1}(1-\tau_c/\Delta t)$ so we can neglect the second term for $\Delta t\gg\tau_c$ and write
\begin{equation}
	\gamma_{mn}(\Delta t)=\frac{1}{\hbar^2}\int_{-\infty}^{+\infty}
	G(\tau)e^{-i(\omega_m+\omega_n)\tau/2}d\tau,\quad |\omega_{mn}|\Delta t\ll1.
	\label{GmnNonZero}
\end{equation}
Finally, if $|\omega_{mn}|\Delta t\gg1$, we have $\gamma_{mn}(\Delta t)=0$. So, in summary, we have, with , $\tau_c\ll\Delta t\ll\tau_S$
\begin{equation}
	\gamma_{mn}(\Delta t)=
	\begin{cases}
		\displaystyle\frac{1}{\hbar^2}\int_{-\infty}^{+\infty}G(\tau)e^{-i(\omega_{m}+\omega_n)\tau/2}d\tau, & |\omega_{mn}|\Delta t\ll1\\[3ex]
		0, & |\omega_{mn}|\Delta t\gg 1
	\end{cases}
	\label{Gmn}
\end{equation}
\end{widetext}
In particular we note that if $m$ and $n$ are of opposite signs, then $|\omega_{mn}|\sim\omega_S$, so if $\Delta t$ is chosen such that $\Delta t\gg\omega_S^{-1}$ it follows that $\gamma_{mn}=0$ for $m$ and $n$ of opposite signs. This tells us that there will be no contributions to the master equation of the form $X_m\rho X_n^\dagger$ where $m$ and $n$ are of opposite signs. For instance, a term such as $X_n\bar{\rho}X_{-n}^\dagger$ will, in the interaction picture, oscillate rapidly as $\exp(2i\omega_nt)$, and are hence averaged over to zero here. Thus eliminated are all terms that are eliminated in the secular approximation as it is usually applied.

These results are insensitive to the choice of $\Delta t$, provided the inequalities in Eq.\ (\ref{Gmn}) are satisfied. If $\Delta t\gg|\omega_{mn}|^{-1}$, the \emph{dissipative} dynamics (i.e., not the system dynamics) have been smoothed over oscillations of frequency $\omega_{mn}$, and if $\Delta t\ll|\omega_{mn}|^{-1}$, there is no smoothing of oscillations of frequency $\omega_{mn}$. For $\omega_{mn}$ values that do not satisfy the inequalities in Eq.\ (\ref{Gmn}), $\gamma_{mn}$ will still be perhaps strongly dependent on the value of $\Delta t$. The interesting case however, is if, for a given system, the values of $\omega_{mn}$ are such that a value of $\Delta t$ can be chosen to satisfy \emph{either} $|\omega_{mn}|^{-1}\ll\Delta t\ll\tau_S$ for all $\omega_{mn}$, the usual (full) secular approximation limit, \emph{or} $\tau_c\ll\Delta t\ll|\omega_{mn}|^{-1}\ll\tau_S$ for some $\omega_{mn}$, dubbed the partial secular approximation below, where it is discussed further.

\subsection{Full secular approximation\label{Section:FSA}}

The secular approximation as it is usually defined requires that the coarse-graining time interval $\Delta t$ be chosen so that $\Delta t\gg|\omega_{mn}|^{-1}$ for all frequency differences $\omega_{mn}=\omega_m-\omega_n$. As mentioned above, as the frequencies $\omega_m$ can be both positive and negative, this condition also includes the requirement that $\Delta t\gg|\omega_m|^{-1}$ for all $\omega_m$, generically written as $\Delta t\gg \omega_S^{-1}$. In that case, the only terms that will survive are those for which $\omega_m=\omega_n$, i.e.\ only the diagonal terms $\gamma_{nn}(\Delta t)$ will be non-zero and the master equation Eq.\ (\ref{TheBXMasterEquation}) will look like
\begin{equation}
	\frac{d\rho}{dt}=-\frac{i}{\hbar}\left[H_S,\rho\right]+\sum_{n}^{}\gamma_{nn}\left[X_n\rho X_n^\dagger,-\tfrac{1}{2}\left\{X_n^\dagger X_n,\rho\right\}\right].
\end{equation}
If we make use of the expression for $G(\tau)$ from Eq.\ (\ref{Gkernel}):
\begin{equation}
	G(\tau)=\int_{0}^{\infty}g(\omega)^2\left[\coth(\hbar\omega/2kT)
	\cos\omega\tau-i\sin\omega\tau\right]d\omega
\end{equation}
so, with $\gamma(\omega)=2\pi g(\omega)^2/\hbar^2$
and $\coth(\hbar\omega/2kT)=2n(\omega)+1$ we get
\begin{equation}
	\gamma_{nn}(\Delta t)=\gamma(|\omega_n|)\left(n(|\omega_n|)+\theta(-\omega_n)\right)
\end{equation}
where $\theta(x)$ is the unit step function. Typically we will ignore the dependence of $\gamma(\omega)$ on $\omega$ for the typical range of values of $\omega_n$, i.e., set $\gamma(\omega)=\gamma$ and we have
\begin{equation}
	\gamma_{nn}(\Delta t)=
		\gamma n(|\omega_n|)+\gamma\theta(-\omega_n)
		\label{Gnn}
\end{equation}
which are the usual results for energy level damping. For such a master equation, with $\gamma_{nn}$ as given by Eq.\ (\ref{Gnn}), the steady state is known to be the expected Gibbs distribution \cite{Mori08}
\begin{equation}
	\rho(\infty)=e^{-H_S/kT}/Z.
\end{equation}
\subsection{Partial secular approximation}

For certain systems, the possibility exists of making more than one choice of $\Delta t$, each still consistent with the constraints of either $|\omega_{mn}|^{-1}\gg$ or $\ll\Delta t$ of Eq.~(\ref{Gmn}). This arises if the frequency differences $\omega_{mn}$ break up into blocks of widely different frequencies, and will occur in cases in which the system can itself be considered as being made up of  interacting subsystems. The result of such an interaction is to introduce splitting in the energy level structure of the non-interacting systems. For an interaction weak in the sense that this splitting, $\Omega$ say, will be much less than the typical transition frequencies $\omega_S$, there can arise frequency differences $\omega_{mn}\sim\Omega$, thereby introducing a new distinct time scale $\sim\Omega^{-1}$ into the dynamics of the system, this, roughly speaking, being the time scale of exchange of energy between the subsystems. This then suggests the possibility of choosing a coarse-graining time scale $\omega_S^{-1}\ll\Delta t\ll\Omega^{-1}$ which is sufficiently short to `capture', within the master equation, the time dependence associated with this energy exchange dynamics, or $\omega_S^{-1}\ll\Omega^{-1}\ll\Delta t$ which will lead to the secular approximation result of Section \ref{Section:FSA}.

So here we will suppose that the frequency differences $\left\lvert \omega_{mn}\right\rvert$ will be either $\lesssim\Omega$, or else $\gg\Omega$, and note that if $\omega_m$ and $\omega_n$ have opposite signs, then $|\omega_{mn}|\sim\omega_S$, so such differences belong to the second group. This establishes a new time scale with respect to which the coarse-graining can be imposed, and which leads to a master equation which is insensitive to the choice of $\Delta t$, provided any such choice satisfies the earlier stated conditions. Some simple models illustrating the situation will be presented below.

Physically, what is being done here is to smooth the dynamics on a time scale such that any oscillations at a frequency $\Omega$ are either smoothed over, $\Delta t\gg\Omega^{-1}$, which is the secular approximation case, or not, $\Delta t\ll\Omega^{-1}$, which is below referred to as the partial secular approximation.

Given the conditions outlined above satisfied by the frequency differences $\omega_{mn}$, we can now choose $\Delta t$ to be such that $\Delta t\ll|\omega_{mn}|^{-1}$ for all pairs of frequencies $\omega_m,\omega_n$ such that $|\omega_{mn}|\lesssim\Omega$, but $\Delta t\gg|\omega_{mn}|^{-1}$ for all pairs of frequencies $\omega_m,\omega_n$ such that $|\omega_{mn}|\gg\Omega$. In this case, certain of the off-diagonal elements $\gamma_{mn}(\Delta t)$ will be non-zero, and from Eq.~(\ref{Gmn}) and using the expression for $G(\tau)$ in Eq.\ (\ref{Gkernel}), are given by
\begin{multline}
	\gamma_{mn}(\Delta t)=\gamma n(|\omega_m+\omega_n|/2)
	+\gamma\theta(-\omega_m-\omega_n),\\
	|\omega_{mn}|\lesssim\Omega,\quad\omega_S^{-1}\ll\Delta t\ll\Omega^{-1}.
	\label{PartialSecularG}
\end{multline}
Finally, noting that for $m$ and $n$ of opposite sign $\omega_{mn}\sim\omega_S$, and with $\gamma_{mn} = 0$ for $\Delta t\gg|\omega_{mn}|^{-1}\sim\omega_S^{-1}$, the partial secular master equation will take the form
\begin{multline}
	\frac{d\rho}{dt}=-\frac{i}{\hbar}\left[H_S,\rho\right]\\
	+\sum_{m,n}^{}\theta_{mn}\gamma_{mn}\left[X_m\rho X_n^\dagger-\tfrac{1}{2}\left\{X_m^\dagger X_n,\rho\right\}\right]
	\label{PartialSecularGa}
\end{multline}
where $\theta_{mn}=1$ if $mn>0$ and $=0$ if $mn<0$.

\subsubsection{Lindblad form retained}

It might be expected in light of the approximations made leading to Eq.~(\ref{PartialSecularG}) that the positivity of the $\bm{\mathit{\gamma}}$ matrix will no longer be guaranteed, and that the master equation will not be of Lindblad form. However, by virtue of the fact that the occupation number $n(|\omega_m+\omega_n|/2)$ is evaluated at the mean of the two frequencies $\omega_m$ and $\omega_n$, it turns out that positivity is in fact retained. 

Positivity amounts to the requirement that
$$\sum_{m,n}^{}a_m^*\gamma_{mn}(\Delta t)a_n\ge 0$$ for all vectors $\mathbf{a}$. Thus we have to evaluate
\begin{equation}
	\sum_{m,n}^{} a_m^*
	\left[n(|\omega_m+\omega_n|/2)+\theta(-\omega_m-\omega_m)\right]a_n.
\end{equation}
Since, for $m$ and $n$ of opposite sign $\omega_{mn}\sim\omega_S$, and with $\gamma_{mn}=0$ for $\Delta t\gg\omega_S^{-1}$, this condition means that the sum will separate into two contributions where both $m$ and $n$ are positive, and where they are both negative. Thus we can write
\begin{align}
	\sum_{m,n}^{}a_m^*\gamma_{mn}(\Delta t)a_n
	=&\sum_{m,n>0}^{}a_m^*n((\omega_m+\omega_n)/2)a_n\\
	&+\sum_{m,n<0}^{}a_m^*\left[n((\omega_m+\omega_n)/2)+1\right]a_n
\end{align}
Writing
\begin{equation}
	n(\omega)=\frac{1}{e^{\hbar\omega/kT}-1}
	=\sum_{p=0}^{\infty}e^{-(p+1)\hbar\omega/kT}
\end{equation}
\begin{widetext}
we find that
\begin{equation}
	\begin{split}
		\sum_{m,n}^{} a_m^*\gamma_{mn}a_n
		=&~\sum_{p=0}^{\infty}
		\left[\left\rvert\sum_{m>0}^{}a_me^{-(p+1)\hbar\omega_m/2kT}\right\rvert^2
		+\left\lvert\sum_{m<0}^{}a_me^{-p\hbar\omega_m/2kT}\right\rvert^2\right]\ge 0
	\end{split}
	\label{GeneralLindblad}
\end{equation}
\end{widetext}
for any choice of $\mathbf{a}$, and hence we conclude that the matrix $\bm{\mathit{\gamma}}$ is positive. Thus the master equation for the partial secular case, with the elements of the matrix $\bm{\mathit{\gamma}}$ defined by the approximate result Eq.~(\ref{PartialSecularG}), will still be Lindblad.

\subsubsection{Equilibriation vs thermalization}

The presence of the off diagonal elements $\gamma_{mn}$ can result in the coherences of the system steady state not vanishing in the $H_S$ energy basis. Consequently the steady state will not, in general, be the expected Gibbs distribution. While this might be perceived as a thermodynamic failing of the partial secular approximated master equation, it has been pointed out by Suba\c{s}{\i} \emph{et al} \cite{Subasi12} (and see also articles cited therein, \cite{vanHove54,Davies76,Geva00,Fleming12}) that a master equation cannot  necessarily be expected to yield the thermal equilibrium state at infinite time, the latter only emerging in the limit of vanishing system-reservoir interaction strength
\begin{equation}
	\lim_{\gamma\to0}\lim_{t\to\infty}\rho(t)=\frac{e^{-H_S/kT}}{\text{Tr}_S[e^{-H_S/kT}]}
	\label{ThermEquil}
\end{equation}
In an example shown below in Section \ref{DampedCoupledQubits}, in the limit of $\gamma\to0$, it is found that these coherences do vanish, so that in the limit of vanishingly small system-reservoir coupling, the expected Gibbs steady state distribution is regained.

\section{Measurement Interpretation of coarse-grained master equation\label{MeasInterp}}

The above results indicate that different choices of $\Delta t$ will result in different Lindblad master equations. This outcome can be looked on as representing different levels of approximation to the exact master equation and hence to the exact system density operator. However, it can be argued that the different choices of $\Delta t$ represent different outcomes due to a physical choice, that of the temporal resolution of observations made on the reservoir. To arrive at this perspective, we need to write the master equation in terms of jump operators that represent the change in the number of photons in the reservoir.
\begin{widetext}
\subsection{Coarse-grained derivative in terms of increments in total photon number}
First consider the coarse-grained derivative Eq.~(\ref{MyEquation}) written in the interaction picture and ignoring energy shift terms,

\begin{equation}
	\frac{\Delta\bar{\rho}}{\Delta t}=
	-\frac{1}{2\hbar^2}\frac{1}{\Delta t}
	\int_{t}^{t+\Delta t}dt_2\int_{t}^{t+\Delta t}dt_1
	\text{Tr}_R\Big[\big[\bar{V}(t_2),[\bar{V}(t_1),\bar{\rho}(t)
	\otimes\rho_R(0)]\big]\Big].
	\label{MyEquation2}
\end{equation}
Since the reservoir in a thermal state is diagonal in the energy basis, it is straightforward to show that we can replace the interaction $\bar{V}(t)$ by
\begin{equation}
	\bar{V}(t)\to\bar{V}_\phi(t)=\bar{X}(t)\bar{B}_\phi(t)=\int_{0}^{\infty}g(\omega)\left(b(\omega)e^{-i(\omega t+\phi)}
	+b^\dagger(\omega)e^{i(\omega t+\phi)}\right)d\omega
\end{equation}
for any value of $\phi$.
\end{widetext}
If we make use of the Heisenberg equation of motion for total photon number $N$, 
\begin{equation}
	N=\int_{0}^{\infty}b^\dagger(\omega)b(\omega)\,d\omega.
\end{equation}
we find that
\begin{equation}
	\frac{dN}{dt}=\hbar^{-1}XB_{-\pi/2}.
	\label{dNdt}
\end{equation}
and hence, correct to second order in the interaction, we can then write in the Heisenberg picture
\begin{equation}
	\bar{V}(t)\approx V_{-\pi/2}(t)=X(t)B_{-\pi/2}(t)=\hbar\frac{dN(t)}{dt}
\end{equation}
and hence to second order in the system reservoir coupling
\begin{equation}
	\int_{t}^{t+\Delta t}\bar{V}(t_1)dt_1
	=\hbar\left(N(t+\Delta t)-N(t)\right)
	=\hbar\Delta N.
\end{equation}
Thus we find that
\begin{equation}
	\frac{\Delta\bar{\rho}}{\Delta t}=
	-\frac{1}{2\Delta t}\text{Tr}_R
	\left[\left[\Delta N,\left[\Delta N,\bar{\rho}(t)
		\otimes\rho_R(0)\right]\right]\right]
\end{equation}
This equation exposes in a formal manner an already familiar physical picture, namely that provided the reservoir is in a energy-diagonal state, the open system dynamics is driven by the gain and loss by the system of quanta from and to the reservoir. In particular we can also determine the kinds of quantum jumps that the coarse-graining method has introduced. 

The coarse-grained derivative as can be written
\begin{align}
	\bar{\rho}(t+\Delta t)
	=&~\bar{\rho}(t)
	-\frac{i}{\hbar}\left[\bar{H}_c\bar{\rho}
	-\bar{\rho}\bar{H}_c^\dagger\right]\Delta t\notag\\
	&+
	\text{Tr}_R
	\left[\Delta N\bar{\rho}(t)
	\otimes\rho_R(0)\Delta N\right]
	\label{DeltatJump}
\end{align}
The expression for the non-Hermitean Hamiltonian $H_c$ can be read off from Eq.~(\ref{MyEquation2}). It is not analysed here, but can be immediately understood as being the `no-jumps' contribution to the stochastic evolution of the sate of the system. The second term describes a quantum jump, the nature of which will be determined below.

The jump term of Eq.\ (\ref{DeltatJump}) can be rewritten as
\begin{align}
	&\text{Tr}_R\left[\Delta N\bar{\rho}\otimes\rho_R(0)\Delta N\right]\notag\\
	&=\frac{\text{Tr}_R\left[\Delta N\bar{\rho}\otimes\rho_R(0)\Delta N\right]}{\text{Tr}_{SR}\left[\Delta N\bar{\rho}\otimes\rho_R(0)\Delta N\right]}
	P(t+\Delta t,t)\Delta t
\end{align}
where the factor
\begin{equation}
	P(t+\Delta t,t)\Delta t=\frac{\text{Tr}_{SR}\left[\Delta N\bar{\rho}\otimes\rho_R(0)\Delta N\right]}{\Delta t}\Delta t
\end{equation}
can be understood as the probability of a jump occuring in the time interval $(t,t+\Delta t)$ which suggests that within the coarse-grained dynamics, the evolution of the system, by virtue of its interaction with the reservoir, is driven by jump processes which involve the loss or gain of photons to and from the reservoir over the coarse graining time interval $\Delta t$, with these jumps described by
\begin{equation}
	\bar{\rho}(t)\to\bar{\rho}(t+\Delta t)=\frac{\text{Tr}_R\left[\Delta N\bar{\rho}\otimes\rho_R\Delta N\right]}{\text{Tr}_{SR}\left[\Delta N\bar{\rho}\otimes\rho_R\Delta N\right]}.
	\label{NonSelJump}
\end{equation}
This is a non-selective jump in that it can be rewritten in terms of jumps that separately lead to a gain or a loss of a photon from the reservoir. To see this, we can separate $B(t)$ into its positive and negative frequency parts:
\begin{equation}
	B(t)=B^{(+)}(t)+B^{(-)}(t)
\end{equation}
with
\begin{equation}
	B^{(+)}(t)=\int_{0}^{\infty}g(\omega)b(\omega)e^{-i\omega t}d\omega=B^{(-)\dagger}(t).
\end{equation}
We can then write Eq.\ (\ref{dNdt}) as
\begin{equation}
	\frac{dN}{dt}=\frac{dN^{(+)}}{dt}+\frac{dN^{(-)}}{dt}
\end{equation}
so that Eq.\ (\ref{NonSelJump}) is
\begin{align}
	\bar{\rho}(t)\to\bar{\rho}(t+\Delta t)=&~\text{Tr}_R\left[\Delta N^{(+)}\bar{\rho}(t)\otimes\rho_R\Delta N^{(-)}\right]\notag\\
	&+\text{Tr}_R\left[\Delta N^{(-)}\bar{\rho}(t)\otimes\rho_R\Delta N^{(+)}\right]
\end{align}
where the cross terms such as $\text{Tr}_R\left[\Delta N^{(-)}\bar{\rho}\otimes\rho_R\Delta N^{(-)}\right]$ will vanish by virtue of the diagonal nature of $\rho_R(0)$. The first term then represents a jump in which the reservoir loses quanta in the time interval $\Delta t$, the second in which it gains quanta. That this gain or loss takes place over an interval of time suggests that it can be modelled in terms of a measurement process that monitors the change in the number of quanta in the reservoir over the time interval $\Delta t$. This is given in the next Section.

\subsection{Measurement interpretation}

We can give this result a measurement intepretation along the lines of the definition of work proposed by \cite{Roncaglia14}. Assume that at the beginning of each interval $\Delta t$ the reservoir is coupled to an auxiliary system $\mathcal{A}$ spanned by a set of states $\left\{|n\rangle,n\in\mathbbm{Z}\right\}$, initially prepared a state $|0\rangle$ via an entangling interaction $U_I$ such that for a reservoir state an eigenstate of the number operator $N$ with $n$ quanta, $|\phi_n\rangle$ say, then
\begin{equation}
	U_I^\dagger\left(|\phi_n\rangle\otimes|0\rangle\right)
	=|\phi_n\rangle\otimes|-n\rangle.
\end{equation}
The system and reservoir then interact over the time interval $(t,t+\Delta t)$ after which time the inverse entangling operation $U_I$ acts. The state of the auxiliary system is then
\begin{widetext}
	\begin{equation}
	\rho_A=\text{Tr}_{SR}\left[U_I\bar{\mathcal{G}}(t+\Delta t,t)
			\left[U^\dagger_I\rho(t)\otimes\rho_R(0)
			\otimes|0\rangle\langle 0|U_I\right]U_I^\dagger\right].
\end{equation}
\end{widetext}
Introducing projection operators $P_n$ on the Hilbert space of the reservoir such that
\begin{equation}
	P_n|\phi_n\rangle=|\phi_n\rangle,\quad P_mP_n=P_n\delta_{mn},\quad \sum_{m=0}^{\infty}P_m=\mathbbm{1}
\end{equation}
and using $\left[P_m,\rho_R(0)\right]=0$ we find that this becomes
\begin{equation}
	\rho_A=\sum_{m,n}^{}|n-m\rangle\langle n-m|P(n|m)
\end{equation}
where
\begin{equation}
	P(n|m)=\text{Tr}_{SR}\left[P_n\bar{\mathcal{G}}(t+\Delta t,t)
		\left[\rho(t)\otimes\rho_R(0)P_m\right]\right]
\end{equation}
is the probability that $n$ quanta are measured in the reservoir at time $t+\Delta t$ given that $m$ were measured there at time $t$. Thus, the probability that $k$ quanta have been added to the reservoir in this time interval will be
\begin{align}
		P(k)=&~\sum_{m=0}^{\infty}P(m+k|m)\notag\\
		=&~\sum_{m=0}^{\infty}
		\text{Tr}_{SR}\left[\bar{\mathcal{G}}(t+\Delta t,t)
		\left[\rho(t)\otimes\rho_R(0)P_m\right]P_{m+k}\right].
\end{align}
\begin{widetext}
	Substituting for $\bar{\mathcal{G}}(t+\Delta t,t)$ to second order, Eq.\ (\ref{GBar2ndOrder}), we then find
\begin{align}
		P(k)=&~\left(1-\hbar^{-2}\text{Tr}_{SR}\left[\left(\mathcal{V}^{(-)}\mathcal{V}^{(+)}+\mathcal{V}^{(+)}\mathcal{V}^{(-)}\right)\rho(t)\otimes\rho_R(0)\right]\right)\delta_{k0}\notag\\
		&+\hbar^{-2}\text{Tr}_{SR}\left[\mathcal{V}^{(-)}\mathcal{V}^{(+)}\rho(t)\otimes\rho_R(0)\right]\delta_{k1}
		+\hbar^{-2}\text{Tr}_{SR}\left[\mathcal{V}^{(+)}\mathcal{V}^{(-)}\rho(t)\otimes\rho_R(0)\right]\delta_{k,-1}
\end{align}
\end{widetext}
where
\begin{equation}
	\mathcal{V}^{(\pm)}=\int_{0}^{\Delta t}\bar{X}(t)\bar{B}^{(\pm)}(t)dt.
\end{equation}
Thus this probability $P(k)$ breaks up, not unexpectedly, into three contributions, the first where there is no loss or gain by the reservoir and is to be associated with the non-Hermitean term (the `no jump' contribution to the stochastic evolution) of Eq.\ (\ref{DeltatJump}), and not analysed here, while the remaining two terms are the jump contributions assocaited with the reservoir gaining or losing a qanta in the time interval $\Delta t$.

\section{More than one reservoir\label{Section:MoreThanOneReservoir}}

In order to be able to study heat transport between reservoirs of different temperatures, the above formalism has to be generalised to take into account the possibility of the system being in contact with more than one reservoir, these reservoirs not necessarily being at the same temperature. In such a case, the interaction with the reservoirs will take the form
\begin{equation}
	V=\sum_{p}^{}B_pX_p.
\end{equation}
However, provided the reservoirs are independent systems, which means
\begin{equation}
	\left[B_m,B_n\right]=0\quad m\ne n
\end{equation}
and that the reservoirs are all prepared in a thermal state, not necessarily of the same temperature
\begin{equation}
	\rho_{R_p}=\frac{e^{-H_{R_p}/kT_p}}{Z_p}
\end{equation}
so that $\text{Tr}_{R_p}[\rho_{R_p}]=0$, the dissipative part of the master equation Eq.\ (\ref{MyEquation}) will become
\begin{widetext}
\begin{equation}
	\mathcal{D}[\rho]=\sum_{p}^{}\mathcal{D}_p[\rho]=\frac{1}{\hbar^2\Delta t}\sum_{p}^{}
	\int_{0}^{\Delta t}dt_2\int_{0}^{\Delta t}dt_1G_p(t_2-t_1)\left[\bar{X}_p(t_1)\rho\bar{X}_p(t_2)-\tfrac{1}{2}\left\{\bar{X}_p(t_2)\bar{X}_p(t_1),\rho\right\}\right]
	\label{MultiReservoirs}
\end{equation}
\end{widetext}
with
\begin{equation}
	G_p(t_2-t_1)=\text{Tr}_{Rp}\left[\bar{B}_p(t_2)\bar{B}_p(t_1)
	\rho_{R_p}\right]
\end{equation}
which can then be evaluated as for the single reservoir case. Examples of this situation are examined in the following Section.

\section{Illustrative examples\label{Examples}}

Below we will give two examples in which the system has an internal time scale $\Omega^{-1}$. There then arises two choices of how the coarse-graining interval $\Delta t$ can be chosen, either much larger or much smaller that $\Omega^{-1}$, leading to two different master equations for the same system. The differences in the two master equations, and their physical meaning is studied below. Particular attention is paid to the predictions for the full and partial secular approximation results in the case in which $\Omega$ is treated as a parameter that can undergo a quasi-static reduction towards zero. In both cases a global approach is adopted in the sense that the reservoir(s) are coupled to the delocalised energy eigenstates of the composite system.

\subsection{Damped coupled qubits\label{DampedCoupledQubits}}

Consider a system consisting of a pair of interacting qubits, with one or both qubits coupled to a thermal reservoir, as treated in, for instance \cite{Scala08,Wu13,Decordi17,Hofer17}, with the coupling between the qubits being allowed to vary from some initially strong value, to zero. The model could be used, for instance, to study the work done by allowing the coupling between the qubits to decrease to zero (which would be equivalent, e.g., to `separating the qubits' if the coupling is modelled as dipole interaction). The first instance, one isolated and the other coupled to a thermal reservoir, is discussed here; the generalisation to both qubits coupled to separate reservoirs is considered further below.

The qubit pair Hamiltonian is
\begin{equation}
	{H}_S
	=\tfrac{1}{2}\hbar\omega_S\left({\sigma}_{1z}+{\sigma}_{2z}\right)
	+\hbar\Omega\left({\sigma}_{1+}{\sigma}_{2-}
	+{\sigma}_{1-}{\sigma}_{2+}\right).
	\label{2qubitHamiltonian}
\end{equation}
and the eigenstates and eigenvalues of ${H}_S$ are
\begin{equation}
	\begin{alignedat}{2}
		|u\rangle=&|e_1\rangle|e_2\rangle,&\qquad &E_u=\hbar\omega_S\\
		|\pm\rangle=&\frac{1}{\!\sqrt{2}}\left(|e_1\rangle|g_2\rangle
		\pm|g_1\rangle|e_2\rangle\right),&\quad &E_\pm=\pm\hbar\Omega\\
		|l\rangle=&|g_1\rangle|g_2\rangle,&\quad &E_l=-\hbar\omega_S.
	\end{alignedat}
	\label{eigenstates}
\end{equation}
Assuming the coupling to the reservoir is via qubit 2, the interaction is given by
$V={B}{\sigma}_{2x}$ where the reservoir operator $B$ is as defined earlier, Eq.~(\ref{ReservoirOperator}). We require $\sigma_{2x}$ in the interaction picture
\begin{equation}
	\bar{\sigma}_{2x}(t)=e^{i{H}_St/\hbar}{\sigma}_{2x}
	e^{-i{H}_St/\hbar}=\sum_{n=-2}^{2}e^{i\omega_nt}{X}_n
\end{equation}
where the transition frequencies $\omega_n$ and associated operators ${X}_n$ with $\omega_{-n}=-\omega_n$ and $X_n^\dagger=X_{-n}$ are
\begin{equation}
	\begin{alignedat}{4}
		\omega_{1}=&~\omega_S-\Omega, & \qquad &&
		{X}_{1}=&~\frac{1}{\!\sqrt{2}}
		\left(|u\rangle\langle +|-|-\rangle\langle l|\right)\\
		\omega_{2}=&~\omega_S+\Omega, & \qquad &&
		{X}_{2}=&~\frac{1}{\!\sqrt{2}}
		\left(|+\rangle\langle l|+|u\rangle\langle -|\right)
		\label{XDefinitions}
	\end{alignedat}
\end{equation}
and where, for completeness, we could also define ${X}_0=0,\omega_0=0$. These operators can also be written in terms of individual qubit operators as
\begin{equation}
		X_n=\tfrac{1}{2}\left(\mathbbm{1}_1
		\otimes\sigma_{2+}-(-1)^n\sigma_{1+}\otimes\sigma_{2z}\right),\quad n=1,2.
		\label{XInTermsOfSigma}
\end{equation}
The master equation Eq.\ (\ref{TheBXMasterEquation}) will then be determined by the choice of the coarse-graining interval that arises in the calculation of the $\gamma_{mn}$.

\subsubsection{Full secular approximation}

In this case, coarse graining is done on a time scale $\Delta t$ given by $\Delta t\gg\Omega^{-1}$. Only the diagonal elements of the matrix $\bm{\mathit{\gamma}}$ survive and are given by, with $n_\pm=n(\omega_S\pm\Omega)$
\begin{align}
	\bm{\mathit{\gamma}}
	=&~
	\begin{pmatrix}
		\gamma_{-2,-2} & \gamma_{-2,-1} & \ldots \\
		\gamma_{-1,-2} & \gamma_{-1,-1} & \ldots \\
		\vdots & \vdots & \vdots
	\end{pmatrix}\notag\\
	=&~\gamma
	\begin{pmatrix}
		n_++1 & 0 & 0 & 0\\
		0 & n_-+1 & 0 & 0\\
		0 & 0 & n_- & 0\\
		0 & 0 & 0 & n_+
	\end{pmatrix}.
	\label{GDiag}
\end{align}
The master equation is
\begin{align}
		\frac{d{\rho}}{dt}=&~-\frac{i}{\hbar}\left[{H}_S,{\rho}\right]
	+(n_++1)\gamma\left[{X}_2^\dagger{\rho} {X}_2-\tfrac{1}{2}\left\{{X}_2{X}_2^\dagger,{\rho}\right\}\right]\notag\\
	&+(n_-+1)\gamma\left[{X}_1^\dagger{\rho} {X}_1-\tfrac{1}{2}\left\{{X}_1{X}_1^\dagger,{\rho}\right\}\right]\notag\\
	&+n_-\gamma\left[{X}_1{\rho} {X}_1^\dagger-\tfrac{1}{2}\left\{{X}_1^\dagger {X}_1,{\rho}\right\}\right]\notag\\
	&+n_+\gamma\left[{X}_2{\rho} {X}_2^\dagger-\tfrac{1}{2}\left\{{X}_2^\dagger {X}_2,{\rho}\right\}\right].
\label{SecularME}
\end{align}
The equilibrium steady state solution of this can be readily shown to be given by the expected Boltzmann distribution of energy level probabilities for a system in thermal equilibrium at temperature $T$, that is
\begin{equation}
	\rho(\infty)=Z(\Omega)^{-1}e^{-H_S/kT}\label{BoltzmannProbs}
\end{equation}
with the partition function $Z(\Omega)$ given by 
\begin{equation}
		Z(\Omega)=\text{Tr}\left[e^{-{H}_S/kT}\right]
		=2\left[\cosh\left(\hbar\omega_S/kT\right)
		+\cosh\left(\hbar\Omega/kT\right)\right].
	\label{ZOmega}
\end{equation}
The steady state reduced density operators $\rho_n(\infty)$ for each qubit $n=1,2$ readily follows and are given by
\begin{align}
	{\rho}_n(\infty)=&~Z(\Omega)^{-1}\left[\left(e^{-\hbar\omega_S/kT}
	+\cosh(\hbar\Omega/kT)\right)|e_n\rangle\langle e_n|\right.\notag\\
	&+\left.\left(e^{\hbar\omega_S/kT}
	+\cosh(\hbar\Omega/kT)\right)|g_n\rangle\langle g_n|\right].\label{SecularReducedSteadyState}
\end{align}
i.e., the density operators for the two qubits are identical. 

In the spirit of the quantum trajectory formalism, we can consider the kinds of quantum jumps induced by this master equation. It is most easy to see this if we assume that the coupling between the two qubits is weak in the sense that $\Omega\ll\omega_S$. In this case, provided also that $\Omega\ll kT/\hbar$, we can neglect the $\Omega$ dependence of $n_\pm$, whilst retaining the $\Omega$ dependence of the Hamiltonian $H_S$. We find in this limit, with $n_\pm\to n$, that the master equation collapses to
\begin{align}
	\frac{d\rho}{dt}=&~-\frac{i}{\hbar}\left[H_S,\rho\right]\notag\\
	&+\gamma(n+1)\left[\sigma_{2-}\rho\sigma_{2+}
	-\tfrac{1}{2}\left\{\sigma_{2+}\sigma_{2-},\rho\right\}\right]\notag\\
	&+\gamma n\left[\sigma_{2+}\rho\sigma_{2-}
	-\tfrac{1}{2}\left\{\sigma_{2-}\sigma_{2+},\rho\right\}\right]\notag\\
	&+\gamma(n+1)\left[\sigma_{2z}\sigma_{1-}\rho\sigma_{1+}\sigma_{2z}
	-\tfrac{1}{2}\left\{\sigma_{1+}\sigma_{1-},\rho\right\}\right]\notag\\
	&+\gamma n\left[\sigma_{2z}\sigma_{1+}\rho\sigma_{1-}\sigma_{2z}
	-\tfrac{1}{2}\left\{\sigma_{1-}\sigma_{1+},\rho\right\}\right]
	\label{TwoQubitsSecME}
\end{align}
which indicates that both qubits will undergo jumps, i.e., this is still a delocalized master equation.

If this result were to be used to study the behaviour of the system in the limit of the coupling between the two qubits being `turned off', $\Omega\to0$, a difficulty arises. Specifically, for the steady state, we find that the density operator for the combined system factorises:
\begin{equation}
	{\rho}(\infty)={\rho}_1(\infty)\otimes{\rho}_2(\infty)
\end{equation}
with, from Eq.\ (\ref{SecularReducedSteadyState})
\begin{equation}
	{\rho}_n(\infty)=Z(0)^{-1}\left(e^{-\frac{1}{2}\hbar\omega_S/kT}|e_n\rangle\langle e_n|+e^{\frac{1}{2}\hbar\omega_S/kT}|g_n\rangle\langle g_n|\right)
\end{equation}
i.e., both qubits settle into the canonical state for temperature $T$, which is the expected equilibrium state of a pair of non-interacting qubits immersed in a common reservoir of temperature $T$. It is difficult to understand this result if only the second qubit is coupled to the reservoir and there is no interaction between the qubits. The problem is, of course, that the master equation is not valid in this limit. The master equation is derived under the condition that the coarse-graining time interval $\Omega^{-1}\ll\Delta t\ll\tau_S$, which clearly cannot be satisfied as $\Omega$ approaches zero.

\subsubsection{Partial secular approximation}

In this case, coarse graining is on a time scale $\omega_S^{-1}\ll\Delta t\ll\Omega^{-1}$. All the diagonal elements $\gamma_{nn}$ will be as in Eq.~(\ref{GDiag}), while the off-diagonal elements will survive. 
\begin{widetext}
The $\gamma$ matrix becomes
\begin{equation}
	\bm{\mathit{\gamma}}
	=
	\begin{pmatrix}
		\gamma_{-2,-2} & \gamma_{-2,-1} & \ldots \\
		\gamma_{-1,-2} & \gamma_{-1,-1} & \ldots \\
		\vdots & \vdots & \vdots		
	\end{pmatrix}
	=\gamma
	\begin{pmatrix}
		n_++1 & n+1 & 0 & 0\\
		n+1 & n_-+1 & 0 & 0\\
		0 & 0 & n_- & n\\
		0 & 0 & n & n_+
	\end{pmatrix}
\end{equation}
with $n_\pm\equiv n(\omega_S\pm\Omega)$ as before and $n\equiv n(\omega_S)$. The master equation (expanded out in detail in Appendix \ref{Me2Qb}) is given by
	\begin{align}
		\frac{d{\rho}}{dt}
		=-\frac{i}{\hbar}\left[{H}_S,{\rho}\right]
		&~+(n_++1)\gamma\left[{X}_2^\dagger{\rho} 		{X}_2-\tfrac{1}{2}\left\{{X}_2{X}_2^\dagger,{\rho}\right\}\right]+(n+1)\gamma\left[{X}_2^\dagger{\rho} 		{X}_1-\tfrac{1}{2}\left\{{X}_1{X}_2^\dagger,{\rho}\right\}\right]\notag\\
		&~+(n+1)\gamma\left[{X}_1^\dagger{\rho} 		{X}_2-\tfrac{1}{2}\left\{{X}_2{X}_1^\dagger,{\rho}\right\}\right]+(n_-+1)\gamma\left[{X}_1^\dagger{\rho} 		{X}_1-\tfrac{1}{2}\left\{{X}_1{X}_1^\dagger,{\rho}\right\}\right]\notag\\
		&~+n_-\gamma\left[{X}_1{\rho} 		{X}_1^\dagger-\tfrac{1}{2}\left\{{X}_1^\dagger 		{X}_1,{\rho}\right\}\right]+n\gamma\left[{X}_2{\rho} 		{X}_1^\dagger-\tfrac{1}{2}\left\{{X}_1^\dagger 		{X}_2,{\rho}\right\}\right]\notag\\
		&~+n\gamma\left[{X}_1{\rho} 		{X}_2^\dagger-\tfrac{1}{2}\left\{{X}_2^\dagger 		{X}_1,{\rho}\right\}\right]+n_+\gamma\left[{X}_2{\rho} 		{X}_2^\dagger-\tfrac{1}{2}\left\{{X}_2^\dagger 		{X}_2,{\rho}\right\}\right].
	\label{ParSecMasterEqn}
\end{align}
which by the general result Eq.~(\ref{GeneralLindblad}) is a Lindblad master equation. This master equation differs from the secular approximation form by the presence of cross-terms which, in this case, mean that the populations and coherences do not decouple, unlike what is found for the full secular approximation form for the master equation. Some of the consequences of this can be seen in the steady state. The only non-vanishing elements of ${\rho}$ that survive at steady state are the diagonal elements $\rho_{uu}$, $\rho_{++}$, $\rho_{--}$, $\rho_{ll}$, and the off-diagonal elements $\rho_{+-}$ and $\rho_{-+}$ and are given by
	\begin{equation}
	\begin{split}
		\rho_{uu}=&~\frac{e^{-\hbar\omega_S/kT}}{Z(\Omega)}
		+p_0\left(\frac{-1}{4(n_++n_-+1)}+(n+\tfrac{1}{2})\frac{n_++n_-+\tfrac{1}{2}}{(2n_-+1)(2n_++1)}\right)\\
		\rho_{++}=&~\frac{e^{-\hbar\Omega/kT}}{Z(\Omega)}
		+p_0\left(\frac{1}{4(n_++n_-+1)}+(n+\tfrac{1}{2})\frac{n_--n_++\frac{1}{2}}{(2n_++1)(2n_-+1)}\right)\\
		\rho_{--}=&~\frac{e^{\hbar\Omega/kT}}{Z(\Omega)}
		+p_0\left(\frac{1}{4(n_++n_-+1)}+(n+\tfrac{1}{2})\frac{n_+-n_-+\frac{1}{2}}{(2n_++1)(2n_-+1)}\right)\\
		\rho_{ll}=&~\frac{e^{\hbar\omega_S/kT}}{Z(\Omega)}
		+p_0\left(\frac{-1}{4(n_-+n_++1)}-(n+\tfrac{1}{2})\frac{n_++n_-+\frac{3}{2}}{(2n_-+1)(2n_++1)}\right)
	\end{split}
\end{equation}
\end{widetext}
where $Z(\Omega)$ is as in Eq.~(\ref{ZOmega}), and
\begin{align}
	p_0=&~\rho_{+-}+\rho_{-+}
	=\frac{1-\frac{(2n+1)(n_++n_-+1)}{(2n_-+1)(2n_++1)}}{R-(2n+1)^2
	\frac{n_++n_-+1}{(2n_-+1)(2n_++1)}}\notag\\
	R=&~\frac{16\Omega^2}{(n_-+n_++1)\gamma^2}+n_++n_-+1\label{p0}
\end{align}
which, unfortunately, does not seem to allow itself to be simplified any further, and
\begin{equation}
	\rho_{+-}-\rho_{-+}=\frac{-4i\Omega}{(n_++n_-+1)\gamma}(\rho_{+-}+\rho_{-+}).
\end{equation}
So, along with the usual Boltzmann distribution contribution to the populations of the eigenstates of ${H}_S$, there is a contribution in each case due to the coherence term $p_0$, so this result is clearly not a canonical thermal state for the combined qubits. However, in the limit of vanishingly small system-reservoir coupling, $\gamma\to0$, we find that $p_0\to 0$ and we have the same result as earlier, Eq.~(\ref{BoltzmannProbs}), the canonical Boltzmann probabilities, as expected from the general result Eq.\ (\ref{ThermEquil}).

What is of interest is the reduced density operators for the two qubits. We saw earlier that these reduced states were identical for the secular form of the master equation, Eq.~(\ref{SecularME}), but that that result is not acceptable as it gives the incorrect $\Omega\to0$ limit. Here, the reduced states are
\begin{widetext}
	\begin{equation}
	\begin{split}
		{\rho}_n
		=&~Z(\Omega)^{-1}\left[\left(e^{-\hbar\omega_S/kT}+\cosh(\hbar\Omega/kT)
		+\tfrac{1}{2}p_0\left(Z(0)-(-1)^nZ(\Omega)\right)\right)
		|e_1\rangle\langle e_1|\right.\\
		&\hspace{1.5cm}+\left.\left(e^{\hbar\omega_S/kT}+\cosh(\hbar\Omega/kT)
		-\tfrac{1}{2}p_0\left(Z(0)-(-1)^nZ(\Omega)\right)\right)
		|g_1\rangle\langle g_1|\right],\quad n=1,2
	\end{split}
\end{equation}
\end{widetext}
So the reduced states of the two qubits are clearly not the same, the origin of this being, of course, the fact that the steady state density operator for the combined qubit system is not diagonal in the energy basis. So, to preserve this expected asymmetry between the qubit states, the steady state cannot be diagonal, i.e., it cannot be a Gibbs state. This result is enough to drive home the fact that the partial secular approximation result is closer to the exact density operator result, and that the diagonal density thermal equilibrium result is only achieved in the limit of vanishingly small system-reservoir interaction.

Once again, we can consider the kinds of quantum jumps induced by this master equation. This can be most clearly seen if, as before, we consider the limit of weak coupling between the two qubits, $\Omega\ll\omega_S$. We can then neglect the $\Omega$ dependence of $n_\pm$, whilst retaining the $\Omega$ dependence of the Hamiltonian $H_S$. We find, on using Eq.\ (\ref{XInTermsOfSigma}), the master equation collapses to
\begin{align}
	\frac{d{\rho}}{dt}
	=&~-\frac{i}{\hbar}\left[{H}_S,{\rho}\right]
	+(n+1)\gamma\left[{\sigma}_{2-}
	{\rho}{\sigma}_{2+}
	-\tfrac{1}{2}\left\{{\sigma}_{2+}
	{\sigma}_{2-},{\rho}\right\}\right]\notag\\
	&+n\gamma\left[{\sigma}_{2+}{\rho}{\sigma}_{2-}
	-\tfrac{1}{2}\left\{{\sigma}_{2-}
	{\sigma}_{2+},{\rho}\right\}\right].
	\label{SmallOmega}
\end{align}
so the only quantum jumps taking place are of qubit 2; the dissipative term describes dissipation of qubit 2 only. This is the expected form of a local master equation, i.e., in which the reservoir is coupled solely to the local energy eigenstates of the qubit coupled to the reservoir. This result was not found when the same limit is taken for the full secular approximation form for the master equation.

If we take the limit $\Omega\to0$ in $H_S$ then there is no time independent steady state, in general, unless qubit 1 is initially in a mixture of eigenstates of ${H}_S$. In this case, the result is
\begin{equation}
	\begin{split}
		{\rho}_1
		=&~p_e(0)|e_1\rangle\langle e_1|
		+p_g(0)|g_1\rangle\langle g_1|\\
		{\rho}_2=&~Z_0^{-1}
		\left[e^{-\hbar\omega_S/2kT}|e_2\rangle\langle e_2|
		+e^{\hbar\omega_S/2kT}|g_2\rangle\langle g_2|\right]\\
		Z_0=&~e^{-\hbar\omega_S/2kT}+e^{\hbar\omega_S/2kT}
	\end{split}
\end{equation}
with $\rho={\rho}_1\otimes{\rho}_2$. In other words, the steady state is that in which the qubit in contact with the reservoir ends up in the canonical state, while the qubit outside the reservoir can be found in its initial (diagonal) state.

\subsubsection{Both qubits damped}

An obvious generalization of the above example is the case in which both qubits are damped by separate reservoirs, not necessarily of the same temperature. If qubit 1 is coupled to a reservoir of temperature $T_h$ and qubit 2 to a reservoir of temperature $T_c$, with $T_h\ge T_c$ the required master equations in the secular and partial secular approximations for independent reservoirs (see Section \ref{Section:MoreThanOneReservoir}) follow directly from those given above Eq.\ (\ref{SecularME}) and Eq.\ (\ref{ParSecMasterEqn}) by simply adding on the dissipator associated with qubit 1. This dissipator can be obtained from that for qubit 2 by the simple substitution $|-\rangle\to-|-\rangle$ in the dissipative contributions to the equations for the individual matrix elements. The damping rates will be $\gamma_c$ and $\gamma_h$ and can be assumed to be unequal in general. The previous example then corresponds to the choice of $\gamma_h=0,\gamma_c=\gamma$. The transition frequencies of each qubit are assumed to be identical at $\omega_S$.

Confining our attention to the steady state and for the reservoirs at equal temperatures, we readily find that the full secular approximation form of the master equation predicts a Gibbsian equilibrium state for the system. In contrast, for the partial secular master equation, there is coupling between the system populations and coherences provided that $\gamma_h\ne\gamma_c$, so at steady state there are non-zero coherences in the energy basis, and the populations are not the expected Boltzmann distribution.

For reservoirs at different temperatures, there will be a steady state heat current $J$ between the reservoirs that turn out to be quite different for the secular and partial secular cases. Letting $\mathcal{D}_h[\rho]$ be the dissipators for qubit 1 coupled to the hot reservoir and $\mathcal{D}_c[\rho]$ the dissipator for qubit 2 coupled to the cold reservoir, the heat current between the reservoirs will be given by
\begin{equation}
	J=\text{Tr}_S\left[H_S\left(\mathcal{D}_h-\mathcal{D}_c\right)[\rho]\right].
\end{equation}
We will only concern ourselves with the steady state in the limit of $\Omega\ll\omega_S$ so that, provided also that $\Omega\ll kT/\hbar$, we can replace $n_p(\omega_S\pm\Omega)\approx n_p(\omega_S)$, $p=c,h$. So with 
\begin{equation}
	\begin{aligned}
		\bar{\gamma}=&(\gamma_h+\gamma_c)/2,\\
		\Delta n=&(\gamma_hn_h(\omega_S)-\gamma_cn_c(\omega_S))/\bar{\gamma}\\
		\bar{n}=&(\gamma_hn_h(\omega_S)+\gamma_cn_c(\omega_S))/2\bar{\gamma}\\
		\Delta\gamma=&(\gamma_h-\gamma_c)/\bar{\gamma}
	\end{aligned}
\end{equation}
we find that in the secular approximation case, at steady state, $\rho_{+-}=0$ and $\rho_{--}=\rho_{++}$ and the steady state heat current $J_\text{sec}$ is given by
\begin{equation}
	J_\text{sec}=\bar{\gamma}\hbar\omega_S\left[\Delta n\left(\rho_{ll}-\rho_{uu}\right)-\Delta\gamma(\rho_{uu}+\rho_{++})\right]
\end{equation}
which reduces to
\begin{equation}
	J_{\text{sec}}=\frac{2\gamma_c\gamma_h\hbar\omega_S(n_h-n_c)}{(\gamma_h+\gamma_c)(2\bar{n}+1)}.
\end{equation}
It is clearly the case that the secular approximate case remains at a constant non-zero value as $\Omega\to0$ which is physically unacceptable. The origin of this behaviour can be traced, once again, to the fact that the secular master equation is invalid in the limit of vanishing $\Omega$.

For the partial secular case we find on using the $\Omega\ll\omega_S$ limiting form for the dissipators that, on using Eq.\ (\ref{XInTermsOfSigma}), the master equation collapses to the local form of the master equation
\begin{align}
	\frac{d\rho}{dt}=&-\frac{i}{\hbar}\left[H_S,\rho\right]+(n_c+1)\gamma\left[\sigma_{1-}\rho\sigma_{1+}-\tfrac{1}{2}\left\{\sigma_{1+}\sigma_{1-},\rho\right\}\right]\notag\\
	&+n_c\gamma\left[\sigma_{1+}\rho\sigma_{1-}-\tfrac{1}{2}\left\{\sigma_{1-}\sigma_{1+},\rho\right\}\right]\notag\\
	&+(n_h+1)\gamma\left[\sigma_{2-}\rho\sigma_{2+}-\tfrac{1}{2}\left\{\sigma_{2+}\sigma_{2-},\rho\right\}\right]\notag\\
		&+n_h\gamma\left[\sigma_{2+}\rho\sigma_{2-}-\tfrac{1}{2}\left\{\sigma_{2-}\sigma_{2+},\rho\right\}\right]
\end{align}
\begin{widetext}
and the heat current becomes
\begin{align}
		J_\text{parsec}=\bar{\gamma}\hbar\omega_S\left[\Delta n\left(\rho_{ll}-\rho_{uu}\right)-\tfrac{1}{2}\Delta\gamma(2\rho_{uu}+\rho_{--}+\rho_{++})-(2\bar{n}+1)(\rho_{+-}+\rho_{-+})\right]
\end{align}
Contributions due to non-vanishing coherences $\rho_{+-}$ now appear. This, in addition to the fact that the steady state populations are not the same as in the secular case as the coherences also contribute there, leads to a different result
	\begin{equation}
	J_{\text{parsec}}
	=J_{\text{sec}}\frac{4\Omega^2}{(2n_c+1)(2n_h+1)\gamma_h\gamma_c+\tfrac{1}{4}[(2n_h+1)\gamma_h-(2n_c+1)\gamma_c](\gamma_h-\gamma_c)+4\Omega^2}
\end{equation}
\end{widetext}
in which case the heat current vanishes for $\Omega\to 0$ as ought to be expected. In the limit when $\Omega^2\gg(2n_h+1)(2n_c+1)\gamma_n\gamma_c$, i.e., when the energy separation of the $|+\rangle$ and $|-\rangle$ states is much larger than their linewidths (but still $\ll\omega_S$), the partial secular and full secular heat currents agree.

\subsection{The tunnelling qubit model\label{Section:TunnellingQBM}}

We shall now apply the above formalism to another example in which there is two clear choices for how the coarse-graining can be invoked, leading to two master equations that would appear to have two different physical interpretations. The model is that of a qubit that can tunnel between two potential wells, e.g., as might be realised with a pair of nearby quantum dots. It is further assumed that the two potential wells are immersed in independent thermal reservoirs. These reservoirs are in general not at the same temperature, so that the tunnelling process will enable the transport of energy (heat) from one reservoir to the other. 
 
The system Hamiltonian is
\begin{align}
	H_S=&~H_A\otimes\mathbbm{1}_T+\mathbbm{1}_A\otimes H_T\notag\\
	=&~\frac{1}{2}\hbar\omega_S\sigma_z+\frac{1}{2}\hbar\Omega\Big(|l\rangle\langle r|+|r\rangle\langle l|\Big)
\end{align}
where the energy eigenstates of $H_A$ are $|e\rangle$ and $|g\rangle$, with $\sigma_z=|e\rangle\langle e|-|g\rangle\langle g|$, and where $|l\rangle$ and $|r\rangle$ are the position eigenstates of the atom, at the site of the left hand and right hand potential wells respectively. It will further be assumed that $\Omega\ll\omega_S$ i.e., that the tunnelling rate will be very much slower that the transition frequency of the qubit.

Note that this is not the case of two independent subsystems coming into interaction: there is no coupling between `system' $A$ and `system' $T$, and the notion of a `local' and a `global' form for the master equation becomes ill-defined. Nevertheless it is meaningful to consider the full secular and partial secular limits of the master equation, and to show that once again, the full secular approximation yields results that are invalid when the tunnelling rate $\Omega$ becomes very small.

Setting
\begin{equation}
	|\pm\rangle=\frac{1}{\sqrt{2}}\left(|l\rangle\pm|r\rangle\right)
\end{equation} 
the eigenstates and eigenvalues of $H_S$ are 
\begin{align}
	H_S|e,\pm\rangle=\tfrac{1}{2}\hbar(\omega_S\pm\Omega)|e,\pm\rangle\notag\\
	H_S|g,\pm\rangle=-\tfrac{1}{2}\hbar(\omega_S\mp\Omega)|g,\pm\rangle
	\label{EigenstatesTunnelling}
\end{align}
The interaction with the reservoirs is given by 
\begin{equation}
	V=B_lP_l\sigma_x+B_rP_r\sigma_x=B_lX_l+B_rX_r
\end{equation}
where $P_n=|n\rangle\langle n|$, $n=l,r$ and where $\left[B_l,B_r\right]=0$ i.e., the reservoirs are independent in which case the analysis of Section \ref{Section:MoreThanOneReservoir} carries through, with the dissipation expressed as the sum of dissipators for each reservoir independently.

We require the time dependence in the interaction picture of the system operators $\bar{X}_l(t)=\bar{P}_l(t)\bar{\sigma}_x(t)$ and $\bar{X}_r=\bar{P}_r(t)\bar{\sigma}_x(t)$ expressed in terms of the associated set of eigenoperators, $X_{lm}$ and $X_{rm}$ where
\begin{equation}
	\bar{X}_{p}(t)=\sum_{m}^{}X_{pm}e^{i\omega_mt},\quad p=l,r.
\end{equation}
We shall do this for $\bar{P}_l(t)\bar{\sigma}_x(t)$, the other following by inspection. If we now put $\Sigma_+=|+\rangle\langle -|$ with $\quad\Sigma_-=\Sigma_+^\dagger$ we have
\begin{equation}
	\bar{P}_l(t)=\frac{1}{2}\Big[1+e^{i\Omega t}\Sigma_++e^{-i\Omega t}\Sigma_-\Big]
\end{equation}
while $\bar{\sigma}_x(t)=\sigma_-e^{-i\omega_St}+\sigma_+e^{i\omega_St}$ and so
\begin{equation}
	\bar{\sigma}_x(t)\bar{P}_l(t)
	=\sum_{m=-3}^{3}X_{lm}e^{i\omega_m t}
\end{equation}
with the individual elements $X_{lm}$ and frequencies $\omega_m$, with $X_{l,-m}=X^\dagger_{l,m}$ and $\omega_{-m}=-\omega_m$,  given by
\begin{equation}
	\begin{pmatrix}
		X_{l1}\\
		X_{l2}\\
		X_{l3}
	\end{pmatrix}
	=\frac{1}{2}
	\begin{pmatrix}
		\sigma_+\Sigma_- 
		\\ \sigma_+ \\  \sigma_+\Sigma_+
	\end{pmatrix}~~~~~\text{and}~~~~
	\begin{pmatrix}
		\omega_1\\
		\omega_2\\
		\omega_3
	\end{pmatrix}
	=
	\begin{pmatrix}
		\omega_S-\Omega\\
		\omega_S\\
		\omega_S+\Omega
	\end{pmatrix}.
	\label{Xleft}
\end{equation}
with the understanding that $X_{l0}\equiv 0$.

In a similar way, the corresponding result for the coupling to the right hand reservoir follows
\begin{equation}
	\bar{\sigma}_x(t)\bar{P}_r(t)=\sum_{m=-3}^{3}X_{rm}e^{i\omega_mt}
\end{equation}
\begin{widetext}
where
\begin{equation}
	\begin{pmatrix}
		X_{r1}\\
		X_{r2}\\
		X_{r3}
	\end{pmatrix}
	=\frac{1}{2}
	\begin{pmatrix}
		-\sigma_+\Sigma_- 
		\\ \sigma_+ \\  -\sigma_+\Sigma_+
	\end{pmatrix}.
	\label{Xright}
\end{equation}
The dissipator here will then consist of two contributions, as given by Eq.\ (\ref{MultiReservoirs}) with the elements $\gamma_{pmn}(\Delta t)$ given by Eq.~(\ref{PartialSecularG}).

	A matrix of values of $\omega_{mn}=\omega_m-\omega_n$ (excluding the row $\omega_{0n}$ and column $\omega_{m0}$) is useful to get an overview the frequency differences:
\begin{equation}
	\begin{pmatrix}
		\omega_{-3,-3} & \omega_{-3,-2} & \ldots \\
		\omega_{-2,-3} & \omega_{-2,-2} & \ldots \\
		\vdots & \vdots & \vdots\\
	\end{pmatrix}=\begin{pmatrix}
		0 & -\Omega & -2\Omega & -2\omega_S & -2\omega_S-\Omega & -2\omega_S-2\Omega\\
		\Omega & 0 & -\Omega & -2\omega_S+\Omega & -2\omega_S & -2\omega_S-\Omega\\
		2\Omega & \Omega & 0 & -2\omega_S+2\Omega & -2\omega_S+\Omega & -2\omega_S\\
		2\omega_S & 2\omega_S-\Omega & 2\omega_S-2\Omega & 0 & -\Omega & -2\Omega\\
		2\omega_S-\Omega & 2\omega_S & 2\omega_S+\Omega & \Omega & 0 & -\Omega\\
		2\omega_S+2\Omega & 2\omega_S+\Omega & 2\omega_S & 2\Omega & \Omega & 0
	\end{pmatrix}.
	\label{omegamatrix}
\end{equation}
\end{widetext}
For $\Omega\ll\omega_S$ there appears two distinct time scales, $\sim\Omega^{-1}$ and $\sim\omega_S^{-1}$, from which we can construct the possible values for the matrix of $\gamma_{pmn}(\Delta t)$ values, depending on the choice of coarse-graining. Two cases are of interest: $\Delta t\gg\Omega^{-1}$ and $\Omega^{-1}\gg\Delta t\gg\omega_S^{-1}$, the first corresponding to the full secular approximation, the second to a partial secular approximation.

\subsubsection{Full secular approximation}

In the full secular case, $\Delta t\gg\Omega^{-1}$, only the $\gamma_{pnn}(\Delta t)$ are non-zero, i.e., from Eq.~(\ref{Gnn}), 
\begin{equation}
	\gamma_{pnn}(\Delta t)=\gamma\left(n_p(|\omega_n|)+\theta(-\omega_n)\right),\quad p=l,r
\end{equation}
with, for reservoirs at temperatures $T_p$, $p=l,r$
\begin{equation}
	n_p(\omega)=(e^{\hbar\omega/kT_p}-1)^{-1}.
\end{equation}
The master equation becomes
\begin{equation}
	\frac{d\rho}{dt}=-\frac{i}{\hbar}\left[H_S,\rho\right]+\mathcal{D}_l[\rho]+\mathcal{D}_r[\rho]
\end{equation}
where the dissipators are given by
\begin{equation}
	\begin{aligned}
	&~\mathcal{D}_l[\rho]+\mathcal{D}_r[\rho]\\
	&=\tfrac{1}{4}\gamma \left(n_l(\omega_S)+1\right)
	\left[\sigma_-\Sigma_-\rho\Sigma_+\sigma_+
	-\tfrac{1}{2}\left\{\Sigma_+\sigma_+\sigma_-\Sigma_-,\rho\right\}\right]\\
	&~+\tfrac{1}{4}\gamma\left(n_l(\omega_S)+1\right)
	\left[\sigma_-\rho\sigma_+
	-\tfrac{1}{2}\{\sigma_+\sigma_-,\rho\}\right]\\
	&~+\tfrac{1}{4}\gamma\left(n_l(\omega_S-\Omega)+1\right)
	\left[\sigma_-\Sigma_+\rho\Sigma_-\sigma_+
	-\tfrac{1}{2}\left\{\Sigma_-\sigma_+\sigma_-\Sigma_+,\rho\right\}\right]\\
	&~+\tfrac{1}{4}\gamma n_l(\omega_S-\Omega)
	\left[\sigma_+\Sigma_-\rho\Sigma_+\sigma_-
	-\tfrac{1}{2}\left\{\Sigma_+\sigma_-\sigma_+\Sigma_-,\rho\right\}\right]\\
	&~+\tfrac{1}{4}\gamma n_l(\omega_s)
	\left[\sigma_+\rho\sigma_-
	-\tfrac{1}{2}\left\{\sigma_-\sigma_+,\rho\right\}\right]\\
	&~+\tfrac{1}{4}\gamma n_l(\omega_S+\Omega)
	\left[\sigma_+\Sigma_+\rho\Sigma_-\sigma_-
	-\tfrac{1}{2}\left\{\Sigma_-\sigma_-\sigma_+\Sigma_+,\rho\right\}\right]\\
	&~+\left(l\to r,~\Sigma_\pm\to-\Sigma_\pm\right)
	\label{TunnellingDissipator}
\end{aligned}
\end{equation}
These equations do not predict any coupling between the populations and the coherences. The latter damp to zero, leaving a steady state density operator diagonal in the energy basis. In the simplest instance of the two reservoirs being at the same temperature, $T_l=T_r=T$ it is straighforward to show that the steady state population distribution is the expected Boltzmann distribution
\begin{equation}
	\rho_{nn}=e^{-E_n/kT}Z^{-1}, n=1\ldots 4.
\end{equation}
The coarse-graining underlying the derivation of the master equation has its consequences when one considers the quantum trajectory unravelling. In particular, for a jump unravelling, and focusing on the post-jump positional state of the tunnelling qubit when the qubit ends up in its ground state, having emitted a photon into one or the other of the reservoirs, the unnormalised post jump positional state of the qubit will be given by, once again assuming for clarity $\Omega\ll\omega_S$ so that, provided also that $\Omega\ll kT_l/\hbar$, $n_l(\omega_S\pm\Omega)\approx n_l(\omega_S)$ and similarly for $n_r$,
\begin{align}
	&\Sigma_-\rho_{ee}\Sigma_++\rho_{ee}+\Sigma_+\rho_{ee}\Sigma_-\\=&~\left(|-\rangle\langle -|+|+\rangle\langle +|\right)(\rho_{11}+\rho_{22})\notag\\
	&+|+\rangle\langle -|\rho_{12}+|-\rangle\langle +|\rho_{21}
\end{align}
At steady state the coherences $\rho_{12}$ and $\rho_{21}$ will be zero in which case the normalised post jump state is
\begin{equation}
	\tfrac{1}{2}\left(|-\rangle\langle -|+|+\rangle\langle +|\right)
\end{equation}
a mixed state with an equal probability of finding the qubit in either the symmetric or antisymmetric positional states. (There is a slight bias towards the lower energy state $|-\rangle$ if the full frequency dependencies of $n_l$ and $n_r$ are taken into account.) Thus, the jump provides no information concerning the position (either left or right quantum well) of the qubit after the emission has occurred. The tunnelling timescale $\Omega^{-1}$ is much larger than the coarse-graining timescale $\Delta t$, so during the time $\Delta t$ the qubit oscillates many times between the left and right hand reservoirs, so the position of the qubit after the emission occurs cannot be resolved. This can be understood as being due to an uncertainty $\Delta t$ in the instant at which the photon has been emitted into the reservoir.

For unequal temperatures, $T_l\ne T_r$, there is a heat current
\begin{equation}
	J=\text{Tr}_{S}\left[H_S(\mathcal{D}_l-\mathcal{D}_r)[\rho]\right]
	\label{TunnellingHeatCurrent}
\end{equation}
between the reservoirs, mediated by the tunnelling qubit. In the system energy eigenstate basis, Eq.~(\ref{EigenstatesTunnelling}), this current can be shown to be
\begin{widetext}
	\begin{equation}
	\begin{aligned}
		J_\text{secular}
		=&~\tfrac{1}{4}\hbar\gamma\left[(\omega_S+\Omega)\Delta n(\omega_S+\Omega)+\omega_S\Delta n(\omega_S)\right]\left(\rho_{g-,g-}-\rho_{e+,e+}\right)\\
		&~+\tfrac{1}{4}\hbar\gamma\left[\omega_S\Delta n(\omega_S)+(\omega_S-\Omega)\Delta n(\omega_S-\Omega)\right]\left(\rho_{g+,g+}-\rho_{e-,e-}\right).
	\end{aligned}
\end{equation}
\end{widetext}
with $\Delta n(\omega)=n_l(\omega)-n_r(\omega)$. This is a complex expression for arbitrary $\Omega$, but assumes a much simpler form if it is assumed that $\Omega\ll\omega_S$ so that, provided also that $\Omega\ll kT_l/\hbar$, $n_l(\omega_S\pm\Omega)\approx n_l(\omega_S)\equiv n_l$ and similarly for $n_r$. In this case it is found that 
\begin{equation}
	J_{\text{secular}}=\frac{\hbar\omega_S\gamma(n_l-n_r)}{2(n_l+n_r+1)}.
	\label{FUllSecularTunnellingCurrent}
\end{equation}
Of note here is that this does not vanish in the limit of $\Omega\to0$, i.e., there is still a heat current present although there is no tunnelling. This unphysical outcome is a consequence of the fact that the coarse-graining condition $\Delta t\gg\Omega^{-1}$ on which basis the master equation was derived will fail for vanishing $\Omega$. 

\subsubsection{Partial secular approximation}

For the partial secular case $\omega_S^{-1}\ll\Delta t\ll\Omega^{-1}$, from Eq.~(\ref{PartialSecularG}) 
\begin{equation}
	\gamma_{pmn}(\Delta t)
	=\gamma\left(n_p(|\omega_m+\omega_n|/2)+\theta(-\omega_m-\omega_n)\right)
\end{equation}
with $\omega_{m}$  and the $X_{l,m}$ given by Eq.~(\ref{Xleft}) and the $X_{r,m}$ by Eq.~(\ref{Xright}). Substituting into the expression Eq.~(\ref{TunnellingDissipator}) for the dissipator yields a very complex expression presented in Appendix \ref{MeTQb} which
differs from the full secular master equation by the presence of 24 extra terms. Of these terms 16 contribute only if the temperatures of the two reservoirs are not equal. 

If the reservoirs are at the same temperature, in spite of the presence of 8 extra terms in the master equation as compared to the secular master equation, the populations evolve independently of the coherences, and approach the expected canonical Boltzmann distributions at steady state. 

However, the master equation simplifies dramatically under the approximation $n_l(\omega_S\pm\Omega)\approx n_l(\omega_S)$ and similarly for $n_r$, for $\Omega\ll\omega_S$, provided also that $\Omega\ll kT/\hbar$, in which case the master equation reduces to
	\begin{align}
		\frac{d\rho}{dt}=&~-\frac{i}{\hbar}\left[H_S,\rho\right]\notag\\
		&~+\gamma(n_l+1)\left[\sigma_-P_l\rho P_l\sigma_+
		-\tfrac{1}{2}\left\{\sigma_+\sigma_-P_l,\rho\right\}\right]\notag\\
		&~+\gamma n_l\left[\sigma_+P_l\rho P_l\sigma_-
		-\tfrac{1}{2}\left\{\sigma_-\sigma_+P_l,\rho\right\}\right]+(l\to r).
	\end{align}
A quantum jump interpretation of this equation follows from extracting the jump terms from this master equation, given by
\begin{equation}
	\gamma(n_l+1)\sigma_-P_l\rho P_l\sigma_+\gamma n_l\sigma_+P_l\rho P_l\sigma_-+(l\to r).
\end{equation}
The projection operators $P_l$ and $P_r$ clearly show that for a jump in the qubit state accompanied by emission or absorption will also project the qubit into either the left or the right potential well. The coarse-graining time interval $\Delta t$ is now much shorter than the tunnelling time, so that when a jump occurs, the temporal resolution is such that the position of the qubit after the jump can be resolved as being either on the left or the right. This is to be contrasted with what was seen in the secular approximation case, where the temporal resolution implied by the choice of $\Delta t$ cannot resolve the qubits position.

Further consequences of the choice of coarse-graining timescales can be seen by examining the heat current $J$, given by Eq.\ (\ref{TunnellingHeatCurrent}). For the partial secular master equation the steady state current is 
\begin{widetext}
	\begin{align}
		J_\text{parsec}=&~\tfrac{1}{4}\hbar\gamma\left[(\omega_S+\Omega)\Delta n(\omega_S+\Omega)+\omega_S\Delta n(\omega_S)\right]\left(\rho_{g-,g-}-\rho_{e+,e+}\right)\notag\\
		&~+\tfrac{1}{4}\hbar\gamma\left[\omega_S\Delta n(\omega_S)+(\omega_S-\Omega)\Delta n(\omega_S-\Omega)\right]\left(\rho_{g+,g+}-\rho_{e-,e-}\right)\notag\\
		&~+\hbar\gamma\left[\left(\omega_S+\tfrac{1}{2}\Omega\right)\bar{n}(\omega_S+\tfrac{1}{2}\Omega)+(\omega_S-\tfrac{1}{2}\Omega)\bar{n}(\omega_S-\tfrac{1}{2}\Omega)\right]\text{Re}[\rho_{g+,g-}]\notag\\
		&~-\hbar\gamma\left[\left(\omega_S+\tfrac{1}{2}\Omega\right)(\bar{n}(\omega_S+\tfrac{1}{2}\Omega)+1)+(\omega_S-\tfrac{1}{2}\Omega)(\bar{n}(\omega_S-\tfrac{1}{2}\Omega)+1)\right]\text{Re}[\rho_{e+,e-}]
	\end{align}
\end{widetext}
where we note that coherences in the energy basis now contribute. This is a complex expression for arbitrary $\Omega$, but assumes a much simpler form if it is assumed that $\Omega\ll\omega_S$ in which case this expression becomes
\begin{equation}
	J_{\text{parsec}}=J_\text{secular}
	\frac{\tilde{\Omega}^2}{1+(2n_l+1)(2n_r+1)+\tilde{\Omega}^2}
\end{equation}
which, in sharp contrast to the full secular result, Eq.\ (\ref{FUllSecularTunnellingCurrent}), vanishes as the tunneling rate $\Omega\to 0$. Note that the secular and partial secular results come into agreement when the tunnelling rate becomes large.

\section{Conclusions}
The derivation of the master equation by a coarse-grained approach was revisited with attention focussed on firstly the conditions for coarse-graining to be applied, and secondly the importance of timescales in determining the possible forms of the master equation. The consequences of this are observed in the case of a composite system with a well-defined internal time scale that is damped by coupling with one or more reservoirs. This is the scenario of on-going interest in studying the global versus local approaches for deriving master equations for such systems. Difficulties with the global secular approximation form found by others, in particular the erroneus prediction of heat currents in circumstances when none should arise, are reproduced here, but are shown to be resolvable by a change in the choice of coarse-graining timescale. It was also argued that the choice of timescales can be given a measurement interpretation that can be directly related under some circumstances to the time resolution of the kinds of quantum jumps that are predicted by the different master equations.

\acknowledgments{}
The authors wish to acknowledge useful conversations with Steve Barnett, Sarah Croke, Alexei Gilchrist, and Thomas Guff.
\bibliographystyle{apsrev4-1.bst}
\bibliography{JDCBib.bib}

\appendix
\section{Estimation of timescales\label{TimeScales}}

We wish to provide estimates for the correction terms arising in the coarse-graining result Eq.\ (\ref{deltarho}) and in the calculation of $\gamma_{mn}(\Delta t)$ in Eq.\ (\ref{GmnApprox}). We first note that in the integral
\begin{equation}
	\frac{1}{\hbar^2}\int_{-\infty}^{\infty}
		G(\tau)e^{i(\omega_m+\omega_n)\tau/2}d\tau.
\end{equation}
as $G(\tau)$ is assumed to decay on a timescale $\tau_c\ll\omega_S^{-1}$, we can approximate this by
\begin{equation}
	\frac{1}{\hbar^2}\int_{-\infty}^{\infty}
		G(\tau)e^{i(\omega_m+\omega_n)\tau/2}d\tau\approx\frac{1}{\hbar^2}\int_{-\infty}^{\infty}
		G(\tau)d\tau\sim\tau_S^{-1}
\end{equation}
Similarly we can write
\begin{multline}
	\frac{1}{\hbar^2}\frac{1}{\Delta t}\int_{0}^{\infty}
		\left[G(-\tau)e^{i(\omega_m+\omega_n)\tau/2}+G(\tau)e^{-i(\omega_m+\omega_n)\tau/2}\right]\tau d\tau\\
		\approx\frac{1}{\hbar^2}\frac{1}{\Delta t}\int_{0}^{\infty}\left[G(-\tau)+G(\tau)\right]\tau d\tau
		\label{approx1}
\end{multline}
Normalising this expression by
\begin{equation}
	\frac{1}{\hbar^2}\frac{1}{\Delta t}\int_{0}^{\infty}\left[G(-\tau)+G(\tau)\right]d\tau=(\Delta t\,\tau_S)^{-1}
\end{equation}
we have
\begin{multline}
	\frac{1}{\hbar^2}\frac{1}{\Delta t}\int_{0}^{\infty}
		\left[G(-\tau)e^{i(\omega_m+\omega_n)\tau/2}+G(\tau)e^{-i(\omega_m+\omega_n)\tau/2}\right]\tau d\tau\\
		\approx
		\tau_S^{-1}\frac{\int_{0}^{\infty}\left[G(-\tau)+G(\tau)\right]\tau d\tau}{\int_{0}^{\infty}\left[G(-\tau)+G(\tau)\right]d\tau}
		\sim\frac{\tau_c}{\tau_S}
		\label{ratio}
\end{multline}
where we have taken the ratio of integrals in Eq.\ (\ref{ratio}) as an estimate of the temporal width $\tau_c$ of the correlation function $G(t)$, with this result now
leading to Eq.\ (\ref{GmnNonZero}).

If we now turn to the correction term in the coarse-grained result Eq.\ (\ref{deltarho}), an interaction of the form $V=BX$ as analysed in Section \ref{Coupling} will lead to structures of the form of Eq.\ (\ref{GeneralDissipator}) but with changed time limits on the integrals:
\begin{equation}
	\int_{t}^{t+\Delta t}dt_2\int_{0}^{t}dt_1 G(t_2-t_1)[\bar{X}(t_1)\rho\bar{X}(t_2)-\ldots ]
\end{equation}
With $G(t)$ having a temporal width $\sim\tau_c$, any oscillating factors from $\bar{X}(t)$ in this expression can be replaced by unity, so we effectively have to deal with the integral over $G(t_2-t_1)$ only. A change of variable then gives
\begin{equation}
	\int_{0}^{\Delta t}dt_2\int_{0}^{t}dt_1G(t_2+t_1).
\end{equation}
If we further make use of the fact that $G(t)$ has a width $\sim \tau_c$, and requiring $t,\Delta t$ are both $\gg\tau_c$, we can allow the upper limits of the double integral to approach infinity. We do this here by introducing the Fourier transform $\tilde{G}(\omega)$ of $G(t)$ to enable us to write
\begin{multline}
	\int_{0}^{\Delta t}dt_2\int_{0}^{t}dt_1G(t_2+t_1)\\
	=\lim_{\epsilon\to0}\int_{-\infty}^{+\infty}d\omega\tilde{G}(\omega)\left(\int_{0}^{\infty}dte^{-i\omega t-\epsilon t}dt\right)^2\\
	=\int_{0}^{\infty}G(\tau)\tau d\tau\sim\frac{\tau_c}{\tau_S}
\end{multline}
So the correction term in Eq.\ (\ref{deltarho}) will be $\sim\tau_c/(\Delta t\,\tau_S)$ which leads to this term being negligible compared to the first term in that expression (of order $\tau_S^{-1}$) for $\Delta t\gg\tau_c$.

\begin{widetext}
\section{Partial secular master equation for two qubit model\label{Me2Qb}}

The partial secular master equation for the two qubit model studied in Section \ref{DampedCoupledQubits} for each qubit coupled to separate reservoirs in general at different temperatures $T_h$ and $T_c$, and with different damping rates $\gamma_h$ and $\gamma_c$ is given by
\begin{equation}
	\frac{d\rho}{dt}=-i\left[\tfrac{1}{2}\omega_S(\sigma_{z1}+\sigma_{z2})+\Omega(\sigma_{1+}\sigma_{2-}+\sigma_{1-}\sigma_{2+})\right]+\mathcal{D}_\text{sec}[\rho]+\mathcal{D}_\text{nonsec}[\rho]
\end{equation}
where $\mathcal{D}_\text{sec}$ are contributions that appear only in the secular approximation form of the master equation, while $\mathcal{D}_\text{nonsec}$ are further terms that appear in the partial secular form. In terms of the mean decay rate $\bar{\gamma}=\left(\gamma_h+\gamma_c\right)/2$ and using the notation \begin{equation}
	\begin{aligned}
		\bar{n}(\omega)=&(2\bar{\gamma})^{-1}(\gamma_hn_h(\omega)+\gamma_cn_c(\omega))\\
		\Delta n(\omega)=&\bar{\gamma}^{-1}\left(\gamma_hn_h(\omega)-\gamma_cn_c(\omega)\right)
	\end{aligned}
\end{equation}
and defining $L_{ab}=|a\rangle\langle b|$, the secular approximation term $\mathcal{D}_\text{sec}[\rho]$ can be written
\begin{equation}
	\begin{aligned}
		\mathcal{D}_\text{sec}[\rho]=&(\bar{n}(\omega_S+\Omega)+1)\bar{\gamma}\left[L_{l+}\rho L_{+l}+L_{-u}\rho L_{u-}-\tfrac{1}{2}\left\{L_{uu}+L_{++},\rho\right\}\right]\\
		&+(\bar{n}(\omega_S-\Omega)+1)\bar{\gamma}\left[L_{u+}\rho L_{+u}+L_{-l}\rho L_{l-}-\tfrac{1}{2}\left\{L_{++}+L_{ll},\rho\right\}\right]\\
		&+\bar{n}(\omega_S-\Omega)\bar{\gamma}\left[L_{+u}\rho L_{u+}+L_{l-}\rho L_{-l}-\tfrac{1}{2}\left\{L_{--}+L_{uu},\rho\right\}\right]\\
		&+\bar{n}(\omega_S+\Omega)\bar{\gamma}\left[L_{+l}\rho L_{l+}+L_{u-}\rho L_{-u}-\tfrac{1}{2}\left\{L_{--}+L_{ll},\rho\right\}\right]\\
		&+\tfrac{1}{2}\Delta n(\omega_S+\Omega)\bar{\gamma}\left[L_{u+}\rho L_{l-}+L_{-l}\rho L_{+u}-L_{l+}\rho L_{u-}-L_{-u}\rho L_{+l}\right]\\
		&+\tfrac{1}{2}\Delta n(\omega_S-\Omega)\bar{\gamma}\left[L_{+u}\rho L_{-l}+L_{l-}\rho L_{u+}-L_{+l}\rho L_{-u}-L_{u-}\rho L_{l+}\right].
	\end{aligned}
\end{equation}
There is no coupling between the populations and the coherences in these secular contributions. However, this is not the case for the non-secular contributions, given by
\begin{equation}
	\begin{aligned}
		\mathcal{D}_\text{nonsec}[\rho]=&(\bar{n}(\omega_S)+1)\bar{\gamma}\left[L_{l+}\rho L_{u+}+L_{+u}\rho L_{+l}-L_{-u}\rho L_{-l}-L_{l-}\rho L_{u-}\right]\\
		&+\bar{n}(\omega_S)\bar{\gamma}\left[L_{+l}\rho L_{+u}+L_{u+}\rho L_{l+}-L_{u-}\rho L_{l-}-L_{-l}\rho L_{-u}\right]\\
		&+\tfrac{1}{2}\Delta n(\omega_S)\bar{\gamma}\left[L_{l+}\rho L_{-l}+L_{l-}\rho L_{+l}+L_{+l}\rho L_{l-}+L_{-l}\rho L_{l+}\right.\\
		&\left.~~~~~~~~~~~~~~~~~-L_{u+}\rho L_{-u}-L_{u-}\rho L_{+u}-L_{+u}\rho L_{u-}-L_{-u}\rho L_{u+}\right].
	\end{aligned}
\end{equation}
which couple the populations to the coherences $\rho_{+-}$ and $\rho_{-+}$.

The particular case of only one qubit coupled to a reservoir is obtained by setting $\gamma_h=0$ and $\gamma_c=\gamma$.

\section{Partial secular master equation for tunnelling qubit model\label{MeTQb}}

The partial secular master equation for the tunnelling qubit model, with $\bar{n}(\omega)=(n_l(\omega)+n_r(\omega))/2$ and $\Delta n(\omega)=n_l(\omega)-n_r(\omega)$ is presented below, with the dissipator broken into three distinctive contributions:
\begin{equation}
	\frac{d\rho}{dt}=-i\left[\tfrac{1}{2}\omega_S\sigma_z+\tfrac{1}{2}\Omega\left(|l\rangle\langle r|+|r\rangle\langle l|\right),\rho\right]+\mathcal{D}_\text{sec}[\rho]+\mathcal{D}_\text{coh1}[\rho]+\mathcal{D}_\text{coh2}[\rho]
\end{equation}
where the secular contributions to the dissipator are
\begin{align}
	\mathcal{D}_{sec}[\rho]
	=&~\tfrac{1}{2}\gamma\left(\bar{n}(\omega_S+\Omega)+1\right)
	\left[\sigma_-\Sigma_-\rho\Sigma_+\sigma_+
	-\tfrac{1}{2}\left\{\Sigma_+\sigma_+\sigma_-\Sigma_-,\rho\right\}\right]\notag\\
	&~+\tfrac{1}{2}\gamma\left(\bar{n}(\omega_S)+1\right)
	\left[\sigma_-\rho\sigma_+
	-\tfrac{1}{2}\{\sigma_+\sigma_-,\rho\}\right]\notag\\
	&~+\tfrac{1}{2}\gamma\left(\bar{n}(\omega_S-\Omega)+1\right)
	\left[\sigma_-\Sigma_+\rho\Sigma_-\sigma_+
	-\tfrac{1}{2}\left\{\Sigma_-\sigma_+\sigma_-\Sigma_+,\rho\right\}\right]\notag\\
	&~+\tfrac{1}{2}\gamma \bar{n}(\omega_S-\Omega)
	\left[\sigma_+\Sigma_-\rho\Sigma_+\sigma_-
	-\tfrac{1}{2}\left\{\Sigma_+\sigma_-\sigma_+\Sigma_-,\rho\right\}\right]\notag\\
	&~+\tfrac{1}{2}\gamma \bar{n}(\omega_S)\left[\sigma_+\rho\sigma_-
	-\tfrac{1}{2}\left\{\sigma_-\sigma_+,\rho\right\}\right]\notag\\
	&~+\tfrac{1}{2}\gamma \bar{n}(\omega_S+\Omega)\left[\sigma_+\Sigma_+\rho\Sigma_-\sigma_-
	-\tfrac{1}{2}\left\{\Sigma_-\sigma_-\sigma_+\Sigma_+,\rho\right\}\right]
\end{align}
while the partial secular approximation introduces extra contributions that give rise to coherences in the spatial degrees of freedom of the qubit, but do not couple the coherences to the populations,
\begin{align}
	\mathcal{D}_{coh1}=&~+\tfrac{1}{2}\gamma\left(\bar{n}(\omega_S)+1\right)\sigma_-\Sigma_-\rho\Sigma_-\sigma_+
	+\tfrac{1}{2}\gamma\left(\bar{n}(\omega_S)+1\right)\sigma_-\Sigma_+\rho\Sigma_+\sigma_+\notag\\
	&~+\tfrac{1}{2}\gamma \bar{n}(\omega_S)\sigma_+\Sigma_+\rho\Sigma_+\sigma_-
	+\tfrac{1}{2}\gamma \bar{n}(\omega_S)\sigma_+\Sigma_-\rho\Sigma_-\sigma_-\notag\\
	&~+\tfrac{1}{4}\gamma\Delta n(\omega_S+\tfrac{1}{2}\Omega)
	\left[\sigma_-\Sigma_-\rho\sigma_+
	-\tfrac{1}{2}\left\{\sigma_+\sigma_-\Sigma_-,\rho\right\}\right]\notag\\
	&~+\tfrac{1}{4}\gamma\Delta n(\omega_S+\tfrac{1}{2}\Omega)
	\left[\sigma_-\rho\Sigma_+\sigma_+-\tfrac{1}{2}
	\left\{\Sigma_+\sigma_+\sigma_-,\rho\right\}\right]
\end{align}
as well as further terms that also give rise to contributions to the coherences, but are present only if the reservoirs are at different temperatures, in which case coupling between the coherences and the populations does occur:
\begin{align}
	\mathcal{D}_{coh2}=&~\tfrac{1}{4}\gamma\Delta n(\omega_S+\tfrac{1}{2}\Omega)
	\left[\sigma_-\Sigma_-\rho\sigma_+
	-\tfrac{1}{2}\left\{\sigma_+\sigma_-\Sigma_-,\rho\right\}\right] 	+\tfrac{1}{4}\gamma\Delta n(\omega_S+\tfrac{1}{2}\Omega)
	\left[\sigma_-\rho\Sigma_+\sigma_+-\tfrac{1}{2}
	\left\{\Sigma_+\sigma_+\sigma_-,\rho\right\}\right]\notag\\
	&~+\tfrac{1}{4}\gamma \Delta n(\omega_S-\tfrac{1}{2}\Omega)
	\left[\sigma_-\rho\Sigma_-\sigma_+
	-\tfrac{1}{2}\left\{\sigma_+\Sigma_-\sigma_-,\rho\right\}\right]	
	+\tfrac{1}{4}\gamma \Delta n(\omega_S-\tfrac{1}{2}\Omega) 	\left[\sigma_-\Sigma_+\rho\sigma_+
	-\tfrac{1}{2}\left\{\sigma_+\sigma_-\Sigma_+,\rho\right\}\right]\notag\\
	&~+\tfrac{1}{4}\gamma \Delta n(\omega_S-\tfrac{1}{2}\Omega)\left[\sigma_+\rho\Sigma_+\sigma_-
	-\tfrac{1}{2}\left\{\Sigma_+\sigma_-\sigma_+,\rho\right\}\right]
	+\tfrac{1}{4}\gamma \Delta n(\omega_S-\tfrac{1}{2}\Omega)\left[\sigma_+\Sigma_-\rho\sigma_-
	-\tfrac{1}{2}\left\{\sigma_-\sigma_+\Sigma_-,\rho\right\}\right]\notag\\
	&~+\tfrac{1}{4}\gamma \Delta n(\omega_S+\tfrac{1}{2}\Omega)\left[\sigma_+\Sigma_+\rho\sigma_-
	-\tfrac{1}{2}\left\{\sigma_-\sigma_+\Sigma_+,\rho\right\}\right]
	+\tfrac{1}{4}\gamma \Delta n(\omega_S+\tfrac{1}{2}\Omega)\left[\sigma_+\rho\Sigma_-\sigma_-
	-\tfrac{1}{2}\left\{\Sigma_-\sigma_-\sigma_+,\rho\right\}\right].
\end{align}

\end{widetext}
\end{document}